\documentclass[journal]{IEEEtran}
\pdfoutput=1
\usepackage{ifpdf}
\usepackage[pdftex]{graphicx}
\usepackage{amssymb}
\usepackage[nospace,compress]{cite}
\usepackage{array}
\usepackage{stfloats}
\usepackage{xcolor}
\usepackage{enumerate}
\usepackage{algorithm}
\usepackage{algpseudocode}
\usepackage{amsmath}
\usepackage{amsthm}
\usepackage{epstopdf}
\usepackage{float}
\usepackage[detect-all,detect-weight=true]{siunitx} % Package included by Nil.
\usepackage{booktabs}  								% Package included by Nil.
\ifpdf

  \DeclareGraphicsExtensions{.pdf,.jpeg,.png}
 \else

  \DeclareGraphicsExtensions{.eps}
 \fi
\ifCLASSINFOpdf
  
\else
  
\fi

\begin{document}
%\pagenumbering{gobble}

\title{Implicit Cooperative Positioning\\ in Vehicular Networks}

\author{\IEEEauthorblockN{Gloria Soatti, Monica Nicoli, Nil Garcia, Benoit Denis, Ronald Raulefs and Henk Wymeersch}

\thanks{This research was supported in part by \textquotedblleft COPPLAR CampusShuttle cooperative perception \& planning platform\textquotedblright, funded under Strategic Vehicle Research and Innovation Grant No. 2015-04849, by the Horizon2020 project HIGHTS (High precision positioning for cooperative ITS applications) MG-3.5a-2014-636537 and by the project MIE (Mobilit\`{a} Intelligente Ecosostenibile) CTN01\_00034\_594122 funded by the Italian Ministry of Education, University and Research (MIUR) within the framework Cluster Tecnologico Nazionale \textquotedblright Tecnologie per le Smart Communities\textquotedblright.}
\thanks{G. Soatti and M. Nicoli are with the Dipartimento di Elettronica, Informazione e Bioingegneria (DEIB), Politecnico di Milano, 20133 Milano, Italy (e-mail: gloria.soatti@polimi.it; monica.nicoli@polimi.it). Nil Garcia and H. Wymeersch are with the Department of Signals and Systems, Chalmers University of Technology, 41296 Gothenburg, Sweden (e-mail: nilg@chalmers.se; henkw@chalmers.se). B. Denis is with the Laboratory of Electronics and Information Technology (LETI), French Atomic Energy Commission (CEA), 38054 Grenoble Cedex 9, France (email: benoit.denis@cea.fr). R. Raulefs is with the Institute of Communications and Navigation, German Aerospace Center (DLR), 82234 Wessling, Germany (e-mail: ronald.raulefs@dlr.de).}}

\maketitle

\begin{abstract}
Absolute positioning of vehicles is based on Global Navigation Satellite Systems (GNSS) combined with on-board sensors and high-resolution maps. In Cooperative Intelligent Transportation Systems (C-ITS), the positioning performance can be augmented by means of vehicular networks that enable vehicles to share location-related information. This paper presents an Implicit Cooperative Positioning (ICP) algorithm that exploits the Vehicle-to-Vehicle (V2V) connectivity in an innovative manner, avoiding the use of explicit V2V measurements such as ranging. In the ICP approach, vehicles jointly localize non-cooperative physical features (such as people, traffic lights or inactive cars) in the surrounding areas, and use them as common noisy reference points to refine their location estimates. Information on sensed features are fused through V2V links by a consensus procedure, nested within a message passing algorithm, to enhance the vehicle localization accuracy. As positioning does not rely on explicit ranging information between vehicles, the proposed ICP method is amenable to implementation with off-the-shelf vehicular communication hardware. The localization algorithm is validated in different traffic scenarios, including a crossroad area with heterogeneous conditions in terms of feature density and V2V connectivity, as well as a real urban area by using Simulation of Urban MObility (SUMO) for traffic data generation. Performance results show that the proposed ICP method can significantly improve the vehicle location accuracy compared to the stand-alone GNSS, especially in harsh environments, such as in urban canyons, where the GNSS signal is highly degraded or denied.

\end{abstract}


\IEEEpeerreviewmaketitle{}

\section{Introduction}\label{sec:Intro}

Intelligent Transportation Systems (ITS) are becoming a crucial component of our society and precise vehicle positioning is playing a key role in it, such as for assisted or autonomous driving, fleet management, and road safety \cite{PaNi12}. Global Navigation Satellite Systems (GNSS), e.g. Global Positioning System (GPS) or Galileo, have been widely used in ITS. Standard GNSS provide an accuracy of 5-10 meters in open sky areas, i.e., when there is a direct line-of-sight between the vehicle receiver and satellites \cite{kaplan2006}. Augmented by differential corrections and/or multi-constellation receivers, they can achieve meter-level accuracy in ideal operating conditions, or even centimeter level in Real Time Kinematics (RTK) variants \cite{skoglund2016}. However, RTK is still subject to long and unpredictable convergence times when cold starting. Moreover, in urban canyons, the availability of both standard and advanced GNSS systems is limited by adverse local environmental conditions, as GNSS signals can be substantially degraded or mostly blocked. A possible way to cope with the degradation of GNSS signals is to perform graph-based Simultaneous Localization And Mapping (SLAM) \cite{thrun2006graph}. Vehicles localize themselves by building maps of the surrounding environment and fusing the available GNSS information into the mapping process. However, in these approaches vehicles are considered as autonomous entities and not connected with each other, while cooperation through vehicular networking could provide significant benefits, particularly in the context of cooperative autonomous driving applications \cite{shladover2006,During_2016}.

In recent years, there has been growing interest on Cooperative ITS (C-ITS) \cite{Wymeersch-CITS}, where vehicles are able to share information with other vehicles and/or the network infrastructure (e.g., base stations or Road Side Units, RSUs) through respectively Vehicle-to-Vehicle (V2V) and/or Vehicle-to-Infrastructure (V2I) communications. Currently, Vehicle-to-anything (V2X) communications can be achieved by using the available IEEE 802.11p technology, which is a basis of the Dedicated Short Range Communication (DSRC) \cite{US_std} and ITS-G5 standards \cite{EUROPE_std}, in the USA and Europe, respectively. Even more recently, cooperative communications have been considered to improve the positioning availability, integrity and continuity in the specific vehicular context\cite{parker2007,parker2007_2,yao2011,drawil2010,Denis2015,Hoang_WPNC16,ETT_RR13,lee2012,rohani2014}. In this context, RSUs can be used for performing vehicle self-localization as they provide wireless connectivity to passing vehicles \cite{wu2013location,wu2015online,ou2014roadside}. In a first stage, a vehicle retrieves the number and locations of the nearby RSUs \cite{wu2013location,wu2015online}, and then positions itself by measuring two-way time-of-arrivals \cite{ou2014roadside}. However, such solutions require populating roads with many RSUs, and thus their use is limited in the short term.  Instead, Cooperative Positioning (CP) approaches \cite{Wymeersch-2009,Meyer_2016,TSIPN-2016}, which benefit from mobile-to-mobile interactions (i.e., in terms of both measurements and exchanged positional information), have been increasingly adopted. They enable to mitigate the shortcomings of GNSS by incorporating additional GNSS-independent information into the positioning problem. Most of these cooperative localization approaches are applied in a more global hybrid data fusion framework \cite{boukerche2008}. The idea is for instance to combine GNSS measurements with auxiliary information, such as ranging measurements, reference points and digital road maps. Several solutions have thus been proposed based on the measurement of inter-vehicle distances \cite{parker2007,parker2007_2,yao2011,drawil2010,Denis2015,Hoang_WPNC16,ETT_RR13,lee2012,rohani2014}. For example, the works in \cite{parker2007,hoang2016_3,yao2011} propose CP algorithms based on V2V radio ranging techniques, assisted with road map and/or vehicle kinematics information. In \cite{ETT_RR13}, the authors develop a strategy for selecting the best V2V links to be used for the purpose of CP. Another CP approach is proposed in \cite{lee2012}, which relies on Radio-Frequency IDentification (RFID) tags installed along the street to compute the GNSS bias for positioning correction. A CP variant worth mentioning is also cooperative map matching \cite{rohani2014} where GNSS raw measurements and precise road-map information are exchanged between the vehicles to further mitigate GNSS errors. Finally, a consensus-based method has been proposed in \cite{TSIPN-2016} to enable the localization of an entire fleet of entities (i.e., vehicles, pedestrians or any objects), at each member of the fleet, by means of local ranging measurements and repeated device-to-device iterations.

Overall, although the current state-of-the-art approaches have been shown to effectively improve the vehicle positioning accuracy, they either rely on high-complexity techniques, or require dedicated hardware or large-scale infrastructure. In addition, most of them need the vehicle to extract explicit range measurements (e.g., round trip time, time of flight or Received Signal Strength -- RSS) from the radio signals exchanged with neighboring mobile entities. Such measurements tend to be of low quality (e.g., RSS measurements) or are incompatible with IEEE 802.11p (e.g., time-based range measurements that rely on unicast). On the one hand, dedicated wireless ranging technologies (e.g., Impulse Radio - Ultra Wideband~\cite{Hoang_WPNC16}) indeed require specific acquisition schemes and handshake protocols. Accordingly, they might induce extra latency, as well as additional coordination or synchronization constraints (e.g., local scheduling of ranging packets within a pseudo-coordinated time division access), which are deemed challenging in highly scalable ad-hoc contexts. On the other hand, as IEEE 802.11p relies on broadcast transmissions, accurate explicit measurements from V2V communications are not available. Consequently, the above methods cannot be straightforwardly applied into the current vehicular context. 
%%%%%

Asides, vehicles are nowadays equipped with more and more perceptual sensors to detect objects and physical obstacles in their close vicinity. These sensors have been mostly intended for forward collision warning, lane change assistance, automatic park control, autonomous driving and more recently, high-definition cartography~\cite{DPMuller_2017}. Commercially available RAdio Detection And Ranging (RADAR) devices devoted to automotive applications, which operate in millimeter frequency bands, can typically measure relative distances and azimuth angles with respect to passive targets within accuracies of a few decimeters and a few tenths of degree respectively, at ranges up to 250 m and with refresh periods lower than 80 ms. LIght Detection And Ranging (LIDAR) devices, including rotating 2D laser scanners, can detect points in the plane within accuracies of a few centimeters and a few hundredths of degree, at ranges up to 200 m and with refresh periods lower than 50 ms. Finally, camera-based systems relying on pixel analysis can also achieve centimeter-level ranging accuracy at shorter ranges on the order of 10 m and with refresh rates up to 60 frames per second. Combining range and angle information thus enables unambiguous 2D relative positioning. Even if the three standalone technologies cited above are also subject to limitations (e.g., rainy/foggy/snowy weathers, optical/radio obstructions, cost per unit for laser scanners, etc.), they generally claim much better spatial resolution than active wireless technologies. 
%%%%%

\textit{Original Contributions:} In this paper, a new \textit{implicit} cooperative positioning (ICP) technique is proposed to improve the GNSS-based vehicle positioning by sharing information and enabling cooperation amongst vehicles through V2V communication links, while making use of on-board sensing devices rather than active wireless technologies enabling explicit V2V measurements. In particular, an innovative distributed processing framework is proposed where a set of non-cooperative features (e.g., people, traffic lights, trees, etc.) are used as common noisy reference points that are cooperatively localized by the vehicles and \textit{implicitly} used to enhance the vehicle location accuracy. A distributed Gaussian Message Passing (GMP) algorithm is designed to solve the positioning problem, integrating a Kalman filter to track the vehicle dynamics based on the on-board GNSS measurements. Vehicles gather noisy observations about the Vehicle-to-Feature (V2F) relative locations by their on-board equipment (e.g., RADAR, camera-based detector, etc.). Then, they reach a \textit{consensus} on the features' absolute locations by engaging in a distributed cooperative estimation, which implicitly reflects on a more accurate vehicle positioning. Some preliminary results on this approach have been presented in \cite{ITSC_2017}. With respect to this previous work, here the ICP algorithm is modified to account for position and velocity dymamics for both vehicles and features. Moreover, the ICP method is derived analytically, together with the related performance bounds for the ideal case of all-to-all connectivity, and it is validated in a realistic road scenario. In particular, the overall solution is first validated by simulation in a simulated crossroad scenario for varying levels of V2V/V2F connectivity. Then, the assessment of the proposed ICP algorithm is carried out in a real urban area of Bologna city (Italy) with traffic generated by using Simulation of Urban MObility (SUMO) for mixed environment conditions. Numerical results show that the proposed approach is able to significantly increase the GNSS-based vehicle location accuracy, especially in urban areas with high density of features and cooperative vehicles, compensating the performance degradation that is typically observed in these areas due to multipath and non-line-of-sight. 

\begin{figure}[!t]
 \centering 
 \includegraphics[width=0.98\columnwidth]{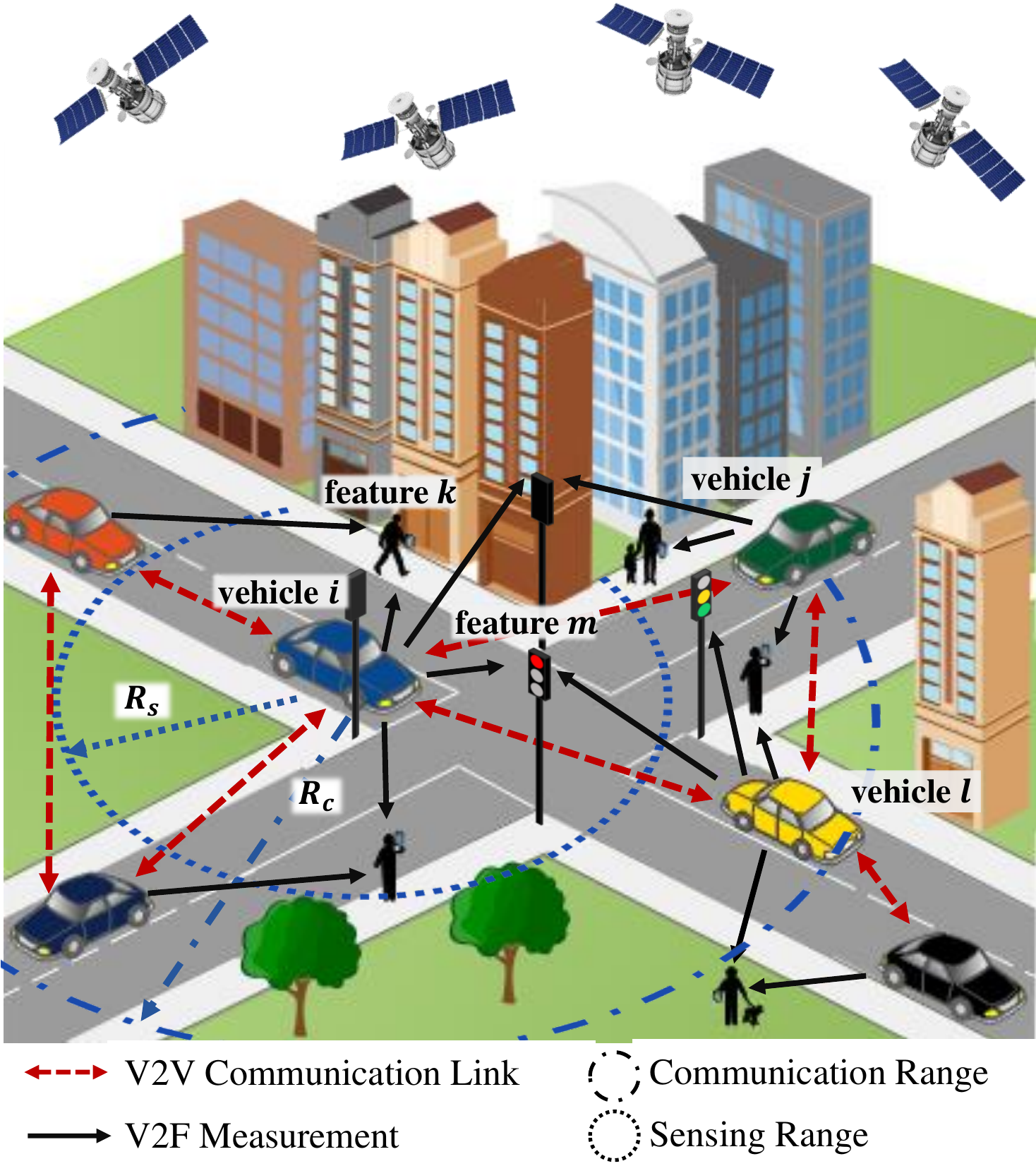}
 \caption{Vehicular network with cooperative vehicles and non-cooperative features.}
 \label{fig:scenario} 
 \end{figure}
 
\section{Problem Formulation and System Model}\label{sec:Prob. fomulation Sistem Mod}

Consider a set of $N_{v}$ interconnected vehicles $\mathcal{V}=\{1,\ldots,N_{v}\}$, deployed over a two-dimensional space as exemplified in Fig. \ref{fig:scenario}. Each vehicle $i\in\mathcal{V}$ has state $\mathbf{x}^{(\mathrm{V})}_{i,t}=[\mathbf{p}_{i,t}^{(\mathrm{V})^\mathrm{T}},\mathbf{v}_{i,t}^{(\mathrm{V})^\mathrm{T}}]^{\mathrm{T}}\in\mathbb{R}^{4\times 1}$, defined as the joint set of the position $\mathbf{p}^{(\mathrm{V})}_{i,t}=[p^{(\mathrm{V})}_{xi,t},p^{(\mathrm{V})}_{yi,t}]^{\mathrm{T}}\in\mathbb{R}^{2\times 1}$ and the velocity $\mathbf{v}^{(\mathrm{V})}_{i,t}=[v^{(\mathrm{V})}_{xi,t},v^{(\mathrm{V})}_{yi,t}]^{\mathrm{T}}\in\mathbb{R}^{2\times 1}$. The state evolves over the time $t$ according to the dynamic model \cite{Gustafsson_2005}: 
\begin{equation}
\mathbf{x}^{(\mathrm{V})}_{i,t}=\mathbf{A}\mathbf{x}^{(\mathrm{V})}_{i,t-1}+\mathbf{Ba}_{i,t-1}+\mathbf{w}^{(\mathrm{V})}_{i,t-1},\label{eq:veh_motion_model}
\end{equation}
with 
\begin{equation}
\mathbf{A}=\left[\begin{array}{cc}
\mathbf{I}_{2} & T_{s}\mathbf{I}_{2}\\
\mathrm{\mathbf{0}_{2\times 2}} & \mathbf{I}_{2}
\end{array}\right],\ \mathbf{B}=\left[\begin{array}{c}
\frac{T_{s}^{2}}{2}\mathbf{I}_{2}\\
T_{s}\mathbf{I}_{2}
\end{array}\right],
\end{equation}
where $\mathbf{A}$ denotes the transition matrix, $\mathbf{B}$ the
matrix relating the vehicle state to the acceleration input $\mathbf{a}_{i,t-1}\in\mathbb{R}^{2\times1}$ here assumed as known (e.g., from an accelerometer), $T_{s}$ is the sampling interval and $\mathbf{w}^{(\mathrm{V})}_{i,t-1}\sim\mathcal{N}(\mathbf{0},\mathbf{Q}^{(\mathrm{V})}_{i,t-1})$
the zero-mean Gaussian driving noise. 

The vehicular network is modelled as a time-varying connected undirected
graph, $\mathcal{G}_{t}=(\mathcal{V},\mathcal{E}_{t})$, with vertices
$\mathcal{V}$ representing the vehicles and edges $\mathcal{E}_{t}$
the V2V communication links. Assuming the communication range equal
to $R_{c}$ at each vehicle, the edge set is $\mathcal{E}_{t}=\{(i,j)\in\mathcal{V}\times\mathcal{V}:||\mathbf{p}^{(\mathrm{V})}_{i,t}-\mathbf{p}^{(\mathrm{V})}_{j,t}||\leq R_{c}\}$,
with vehicles $i$ and $j$ connected if and only if their distance
is lower or equal to $R_{c}$ (as illustrated by dashed red links in Fig.
\ref{fig:scenario}) and with $||\cdot||$ denoting the Frobenius norm. The set of neighbors that directly communicates
with vehicle $i\in\mathcal{V}$ is denoted as $\mathcal{J}_{i,t}=\{j\in\mathcal{V}:(i,j)\in\mathcal{E}_{t}\}$,
its cardinality (i.e., the vehicle degree) as $d_{i,t}=\left\vert \mathcal{J}_{i,t}\right\vert $
and the maximum degree (over all vehicles) as $\Delta_{t}=\max d_{i,t}$.

As illustrated in Fig. \ref{fig:scenario}, the scenario involves
also a set $\mathcal{F}=\{1,\ldots,N_{f}\}$ of $N_{f}$ features
(e.g., people, traffic lights, trees, etc.), either static or mobile, which are non-cooperative entities sensed by the vehicle's on-board equipment (e.g., RADAR, LIDAR, camera-based detector, etc.) and used as common noisy reference points for cooperative localization. Note that features are defined as non-cooperative passive objects since they cannot communicate with each other, or with vehicles, and do not perform any computations and measurements. At time $t$, the state of feature $k$ is $\mathbf{x}^{(\mathrm{F})}_{k,t}=[\mathbf{p}_{k,t}^{(\mathrm{F})^\mathrm{T}},\mathbf{v}_{k,t}^{(\mathrm{F})^\mathrm{T}}]^{\mathrm{T}}\in\mathbb{R}^{4\times 1}$, defined as the joint set of the position $\mathbf{p}^{(\mathrm{F})}_{k,t}=[p^{(\mathrm{F})}_{xk,t},p^{(\mathrm{F})}_{yk,t}]^{\mathrm{T}}\in\mathbb{R}^{2\times 1}$ and the velocity $\mathbf{v}^{(\mathrm{F})}_{k,t}=[v^{(\mathrm{F})}_{xk,t},v^{(\mathrm{F})}_{yk,t}]^{\mathrm{T}}\in\mathbb{R}^{2\times 1}$. The feature state evolves according to the first order Markov model:
\begin{equation}
\mathbf{x}^{(\mathrm{F})}_{k,t}=\mathbf{A}\mathbf{x}^{(\mathrm{F})}_{k,t-1}+\mathbf{w}^{(\mathrm{F})}_{k,t-1},\label{eq:fea_motion_model}
\end{equation}
with zero-mean Gaussian driving noise $\mathbf{w}^{(\mathrm{F})}_{k,t-1}\sim\mathcal{N}(\mathbf{0},\mathbf{Q}^{(\mathrm{F})}_{k,t-1})$. Due to the limited sensing range $R_{s}$,
each vehicle $i$ senses only a subset of all features given by $\mathcal{F}_{i,t}=\{k\in\mathcal{F}:||\mathbf{p}^{(\mathrm{F})}_{k,t}-\mathbf{p}^{(\mathrm{V})}_{i,t}||\leq R_{s}\}\subseteq\mathcal{F}$ (see Fig. \ref{fig:scenario}). Note that in this analysis the focus is on passive objects, but the framework can be easily generalized to handle also non-cooperative active features, equipped with transmitting devices or even combined with a subset of active cooperative units, sometimes denoted as beaconing anchors. 

At time instant $t$, the measurements available at each vehicle for localization are the GNSS location fix and the V2F relative location observations gathered for all the features within the sensing range. The GNSS measurement of vehicle $i$ state is: 
\begin{equation}
\boldsymbol{\rho}^{\mathrm{(GNSS)}}_{i,t}=\mathbf{p}^{(\mathrm{V})}_{i,t}+\mathbf{n}^{\mathrm{(GNSS)}}_{i,t}=\mathbf{P}\mathbf{x}^{(\mathrm{V})}_{i,t}+\mathbf{n}^{\mathrm{(GNSS)}}_{i,t},\label{eq:meas_GNSS_y_in}
\end{equation}
where $\mathbf{P}=\left[\mathbf{I}_{2}\ \mathbf{0}_{2\times 2}\right]$
is the $2\times 4$ matrix selecting the vehicle position, and $\mathbf{n}^{\mathrm{(GNSS)}}_{i,t}\sim\mathcal{N}(\mathbf{0},\mathbf{R}^{\mathrm{(GNSS)}}_{i,t})$
is the GNSS measurement error \cite{parker2007,drawil2010}.
The relative location measurement $\boldsymbol{\rho}^{\mathrm{(V2F)}}_{i,k,t}\in\mathbb{R}^{2\times 1}$
made by vehicle $i$ to its nearby feature $k\in\mathcal{F}_{i,t}$
is modelled as: 
\begin{equation}
\boldsymbol{\rho}^{\mathrm{(V2F)}}_{i,k,t}=q(\mathbf{P}\mathbf{x}^{(\mathrm{F})}_{k,t}-\mathbf{P}\mathbf{x}^{(\mathrm{V})}_{i,t})+\mathbf{n}^{\mathrm{(V2F)}}_{i,k,t}=q(\boldsymbol{\delta}_{i,k,t})+\mathbf{n}^{\mathrm{(V2F)}}_{i,k,t},\label{eq:meas_model_z_ik}
\end{equation}
where $\mathbf{n}^{\mathrm{(V2F)}}_{i,k,t}\sim\mathcal{N}(\mathbf{0},\mathbf{R}^{\mathrm{(V2F)}}_{i,k,t})$
is the measurement uncertainty and the deterministic function $q(\boldsymbol{\delta}_{i,k,t})$
models the relation of the observation to the V2F relative position $\boldsymbol{\delta}_{i,k,t}=\mathbf{p}^{(\mathrm{F})}_{k,t}-\mathbf{p}^{(\mathrm{V})}_{i,t}$.
Depending on the class of on-board sensors detecting the feature,
the measurement may represent the V2F range $q(\boldsymbol{\delta}_{i,k,t})=||\boldsymbol{\delta}_{i,k,t}||$,
the angle $q(\boldsymbol{\delta}_{i,k,t})=\angle(\boldsymbol{\delta}_{i,k,t})$, or the relative position $q(\boldsymbol{\delta}_{i,k,t})=\boldsymbol{\delta}_{i,k,t}$.
In this study, we focus on the latest case assuming that both range and angle measurements are available from on-board vehicle RADAR equipment. This choice allows a more compact information representation and deeper insight into the performance behavior (see Sec. \ref{sec:Centr_solution}). 
Nevertheless, the framework is general enough to include other classes of measurements. The measurement errors $\mathbf{n}^{\mathrm{(GNSS)}}_{i,t}$ and $\mathbf{n}^{\mathrm{(V2F)}}_{i,k,t}$
are assumed to be mutually independent \cite{Wymeersch-2009}, and
also independent over vehicles, features and time. Perfect association
between measurements and features is considered as available at each
vehicle. A discussion on the association problem can be found in Sec. \ref{sec:Cooperative Data Assoc.}.

\section{Centralized ICP Method}

\label{sec:Centr_solution}

Let $\mathbf{x}^{(\mathrm{V})}_{t}=[\mathbf{x}^{(\mathrm{V})}_{i,t}]_{i\in\mathcal{V}}\in\mathbb{R}^{4N_{v}\times1}$
and $\mathbf{x}^{(\mathrm{F})}_{t}=[\mathbf{x}^{(\mathrm{F})}_{k,t}]_{k\in\mathcal{F}}\in\mathbb{R}^{4N_{f}\times1}$
be the vectors collecting all vehicles' and features' states at time
$t$, $\boldsymbol{\rho}_{t}^{\mathrm{(GNSS)}}=[\boldsymbol{\rho}_{i,t}^{\mathrm{(GNSS)}}]_{i\in\mathcal{V}}\in\mathbb{R}^{2N_{v}\times1}$
and $\boldsymbol{\rho}_{t}^{\mathrm{(V2F)}}=[\boldsymbol{\rho}_{i,k,t}^{\mathrm{(V2F)}}]_{i\in\mathcal{V},k\in\mathcal{F}_{i,t}}\in\mathbb{R}^{2M\times1}$
the related GNSS and V2F noisy observations collected by the vehicles. V2F measurements are conveniently
re-indexed as $\boldsymbol{\rho}_{m,t}^{\mathrm{(V2F)}}=\boldsymbol{\rho}_{i_{m},k_{m},t}^{\mathrm{(V2F)}}$
with $m\in\mathcal{M}=\{1,\ldots,M\}$ univocally identifying the
measurement made by vehicle $i_{m}\in\mathcal{V}$ to feature $k_{m}\in\mathcal{F}_{i_{m},t}$,
and $M=\sum_{i=1}^{N_{v}}|\mathcal{F}_{i,t}|$ denoting the total number of V2F measurements. In the centralized
ICP approach, all the observations are gathered by a fusion center
to synchronously estimate the overall dynamic state $\boldsymbol{\theta}_{t}=\Big[\mathbf{x}^{(\mathrm{V})^{\mathrm{T}}}_{t},\mathbf{x}^{(\mathrm{F})^{\mathrm{T}}}_{t}\Big]^{\mathrm{T}}\in\mathbb{R}^{4(N_{v}+N_{f})\times1}$. The augmented measurement model is then: 
\begin{equation}
\boldsymbol{\rho}_{t}=\begin{bmatrix}\boldsymbol{\rho}_{t}^{\mathrm{(GNSS)}}\\
\boldsymbol{\rho}_{t}^{\mathrm{(V2F)}}
\end{bmatrix}=\underset{\mathbf{H}_{t}}{\underbrace{\begin{bmatrix}\tilde{\mathbf{P}} & \mathbf{0}_{2N_{v}\times2N_{f}}\\
\mathbf{M}_{v} & \mathbf{M}_{f}
\end{bmatrix}}}\boldsymbol{\theta}_{t}+\underset{\boldsymbol{n}_{t}}{\underbrace{\begin{bmatrix}\boldsymbol{n}_{t}^{\mathrm{(GNSS)}}\\
\boldsymbol{n}_{t}^{\mathrm{(V2F)}}
\end{bmatrix}}},\label{eq: measurement time t}
\end{equation}
where $\mathbf{H}_{t}$ is the matrix of the known regressors with $\tilde{\mathbf{P}}\hspace{-0.05cm}=\hspace{-0.05cm}\mathbf{I}_{N_{v}}\hspace{-0.05cm}\otimes\mathbf{P}$ and $\otimes$ denoting the Kronecker product. The matrix $\mathbf{M}_{v}\hspace{-0.07cm}=\hspace{-0.08cm}\left[\hspace{-0.02cm}\mathbf{M}_{m,i}\hspace{-0.02cm}\right]$ is block-partitioned
into $M\hspace{-0.04cm}\times\hspace{-0.04cm}N_{v}$ blocks of dimensions $2\hspace{-0.02cm}\times \hspace{-0.02cm}4$ defined as:
$\mathbf{M}_{m,i}=-\mathbf{P}$ if $i=i_{m}$, $\mathbf{M}_{m,i}=\mathbf{0}$
otherwise. Similarly, the matrix $\mathbf{M}_{f}=\left[\mathbf{M}_{m,k}\right]$
is block-partitioned into $M\times N_{f}$ blocks of dimensions $2\times 4$
defined as: $\mathbf{M}_{m,k}=\mathbf{P}$ if $k=k_{m}$, $\mathbf{M}_{m,k}=\mathbf{0}$
otherwise. The Gaussian vector $\mathbf{n}_{t}\sim\mathcal{N}(\mathbf{0},\mathbf{R}_{t})$
aggregates the measurement errors $\mathbf{n}_{t}^{\mathrm{(GNSS)}}=[\mathbf{n}_{i,t}^{\mathrm{(GNSS)}}]_{i\in\mathcal{V}}$
and $\mathbf{n}_{t}^{\mathrm{(V2F)}}=[\mathbf{n}_{m,t}^{\mathrm{(V2F)}}]_{m\in\mathcal{M}}$,
with covariance $\mathbf{R}_{t}=\mathrm{blockdiag}(\mathbf{R}_{1,t}^{\mathrm{(GNSS)}},...,\mathbf{R}_{N_{v},t}^{\mathrm{(GNSS)}},\mathbf{R}_{1,t}^{\mathrm{(V2F)}},...,\mathbf{R}_{M,t}^{\mathrm{(V2F)}})$. V2F sensing errors $\mathbf{n}_{m,t}^{\mathrm{(V2F)}}=\mathbf{n}_{i_{m},k_{m},t}^{\mathrm{(V2F)}}$
and related covariance matrices $\mathbf{R}_{m,t}^{\mathrm{(V2F)}}=\mathbf{R}_{i_{m},k_{m},t}^{\mathrm{(V2F)}}$ are re-indexed as previously discussed according to the V2F measurement numbering.

According to Bayesian filtering \cite{Kay}, the Minimum Mean Square
Error (MMSE) estimate of $\boldsymbol{\theta}_{t}$, given all measurements
$\mathbf{\boldsymbol{\rho}}_{1:t}=\{\boldsymbol{\rho}_{1},\ldots,\boldsymbol{\rho}_{t}\}$
up to time $t$, can be computed as: 
\begin{equation}
\hat{\boldsymbol{\theta}}_{t|t}=\begin{bmatrix}\hat{\mathbf{x}}^{(\mathrm{V})}_{t|t}\\[0.2cm]
\hat{\mathbf{x}}^{(\mathrm{F})}_{t|t}
\end{bmatrix}=\int\boldsymbol{\theta}_{t}p(\boldsymbol{\theta}_{t}|\boldsymbol{\rho}_{1:t})d\boldsymbol{\theta}_{t},\label{eq:MMSE estimate}
\end{equation}
from the posterior probability density function (pdf): 
\begin{equation}
p(\boldsymbol{\theta}_{t}|\boldsymbol{\rho}_{1:t}) \propto p(\mathbf{\boldsymbol{\rho}}_{t}|\boldsymbol{\theta}_{t}) \int\hspace{-0.18cm}p(\boldsymbol{\theta}_{t}|\boldsymbol{\theta}_{t-1})p(\boldsymbol{\theta}_{t-1}|\boldsymbol{\rho}_{1:t-1})d\boldsymbol{\theta}_{t-1},\label{eq:cent_optimum}
\end{equation}
where $\propto$ stands for proportionality. Here the likelihood function is $p(\mathbf{\boldsymbol{\rho}}_{t}|\boldsymbol{\theta}_{t})=p(\boldsymbol{\rho}_{t}^{\mathrm{(GNSS)}}|\mathbf{x}^{(\mathrm{V})}_{t})p(\boldsymbol{\rho}_{t}^{\mathrm{(V2F)}}|\boldsymbol{\theta}_{t})$
as measurements are conditionally independent, while the transition
pdf $p(\boldsymbol{\theta}_{t}|\boldsymbol{\theta}_{t-1})=p(\mathbf{x}^{(\mathrm{V})}_{t}|\mathbf{x}^{(\mathrm{V})}_{t-1})p(\mathbf{x}^{(\mathrm{F})}_{t}|\mathbf{x}^{(\mathrm{F})}_{t-1})$ follows from the mobility models (\ref{eq:veh_motion_model}) and (\ref{eq:fea_motion_model}). Based on these assumptions,
and recalling that models are Gaussian and linear, the centralized
ICP estimate reduces to the Kalman filter: 
\begin{equation}
\hat{\boldsymbol{\theta}}_{t|t}=\hat{\boldsymbol{\theta}}_{t|t-1}\hspace{-0.05cm}+\hspace{-0.05cm}\mathbf{C}_{t|t}\mathbf{H}_{t}^{T}\mathbf{R}_{t}^{-1}\left(\boldsymbol{\rho}_{t}-\mathbf{H}_{t}\hat{\boldsymbol{\theta}}_{t|t-1}\right).\label{eq:Kalman_equivalent}
\end{equation}
where $\hat{\boldsymbol{\theta}}_{t|t-1}$ is the predicted state
for all vehicles and features, 
\begin{equation}
\hat{\boldsymbol{\theta}}_{t|t-1}=\begin{bmatrix}\hat{\mathbf{x}}^{(\mathrm{V})}_{t|t-1}\\[0.2cm]
\hat{\mathbf{x}}^{(\mathrm{F})}_{t|t-1}
\end{bmatrix}=\begin{bmatrix}\tilde{\mathbf{A}}\hat{\mathbf{x}}^{(\mathrm{V})}_{t-1|t-1}+\tilde{\mathbf{B}}\mathbf{a}_{t-1}\\[0.2cm]
\bar{\mathbf{A}}\hat{\mathbf{x}}^{(\mathrm{F})}_{t-1|t-1}
\end{bmatrix},\label{eq:theta_prediction}
\end{equation}
\noindent with $\tilde{\mathbf{A}}=\mathbf{I}_{N_{v}}\otimes\mathbf{A}$, $\tilde{\mathbf{B}}=\mathbf{I}_{N_{v}}\otimes\mathbf{B}$, $\mathbf{a}_{t-1}=[\mathbf{a}_{i,t-1}]_{i\in\mathcal{V}}$ according
to (\ref{eq:veh_motion_model}) and $\bar{\mathbf{A}}=\mathbf{I}_{N_{f}}\otimes\mathbf{A}$ based on (\ref{eq:fea_motion_model}). The covariance of the centralized
ICP estimate is: 
\begin{equation}
\mathbf{C}_{t|t}=\mathrm{Cov}\left(\hat{\boldsymbol{\theta}}_{t|t}\right)=\left(\mathbf{C}_{t|t-1}^{-1}+\mathbf{H}_{t}^{T}\mathbf{R}_{t}^{-1}\mathbf{H}_{t}\right)^{-1},\label{eq:Kalman_covariance}
\end{equation}
where $\mathbf{C}_{t|t-1}$ is the covariance of the prediction: 
\begin{equation}
\mathbf{C}_{t|t-1}=\mathrm{Cov}\left(\hat{\boldsymbol{\theta}}_{t|t-1}\right)=\mathrm{blockdiag}(\mathbf{C}^{(\mathrm{V})}_{t|t-1},\mathbf{C}^{(\mathrm{F})}_{t|t-1}),\label{eq:cov_prediction}
\end{equation}
with $\mathbf{C}^{(\mathrm{V})}_{t|t-1}\hspace{-0.45cm}=\hspace{-0.41cm}\mathrm{blockdiag}(\hspace{-0.02cm}\mathbf{C}^{(\mathrm{V})}_{1,t|t-1}\hspace{-0.02cm},\ldots,\hspace{-0.02cm}\mathbf{C}^{(\mathrm{V})}_{N_{v},t|t-1}\hspace{-0.03cm})\hspace{-0.04cm}$ collecting the prior covariances $\mathbf{C}^{(\mathrm{V})}_{i,t|t-1}=\mathbf{A}\mathbf{C}^{(\mathrm{V})}_{i,t-1|t-1}\mathbf{A}^{\mathrm{T}}\hspace{-0.1cm}+\hspace{-0.1cm}\mathbf{Q}^{(\mathrm{V})}_{i,t-1}$
for all vehicles $i\in\mathcal{V}$, and $\mathbf{C}^{(\mathrm{F})}_{t|t-1}=\mathrm{blockdiag}(\mathbf{C}^{(\mathrm{F})}_{1,t|t-1},...,\mathbf{C}^{(\mathrm{F})}_{N_{f},t|t-1})$
the prior covariances $\mathbf{C}^{(\mathrm{F})}_{k,t|t-1}\hspace{-0.1cm}=\mathbf{A}\mathbf{C}^{(\mathrm{F})}_{k,t-1|t-1}\mathbf{A}^{\mathrm{T}} + \mathbf{Q}^{(\mathrm{F})}_{k,t-1}$ for all features $k\in\mathcal{F}$.

\begin{figure}[!t]
\centering \includegraphics[width=0.96\columnwidth]{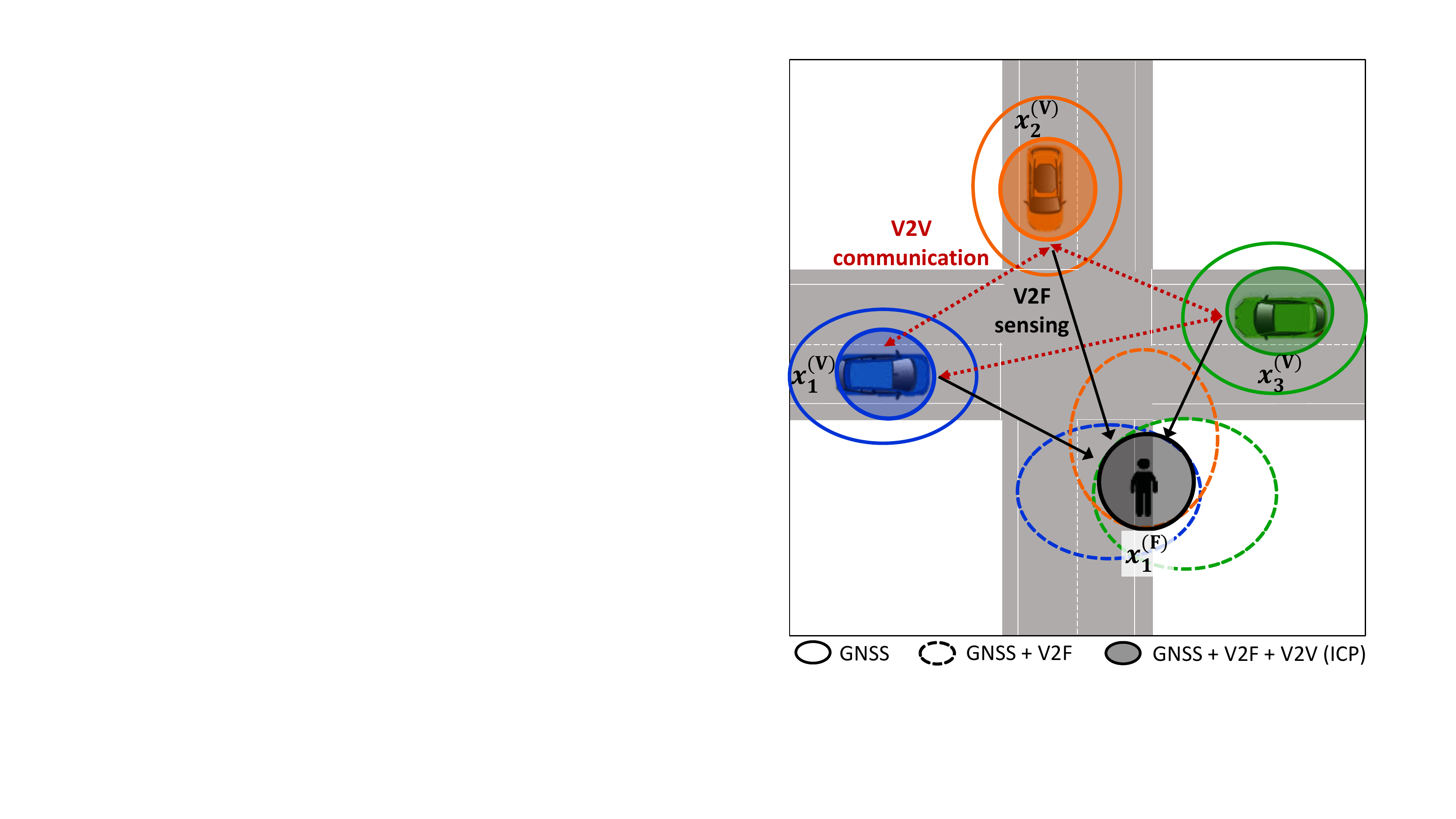}
\caption{ICP example with three vehicles and one passive feature. The location accuracy for the vehicles (solid colored contours) and the feature (dashed colored contours), based on stand-alone GNSS and V2F sensing, is represented as 1-$\sigma$ error ellipse. The accuracy obtained by ICP (colored ellipse) is significantly higher thanks to the cooperative localization of the jointly sensed feature through V2V links.}
\label{fig:ICP_example} 
\end{figure}

\section{Distributed ICP Method}\label{sec:Distrib_algorithm}

The centralized ICP solution is not
practical for large-scale networks: not only does a central computing
unit constitute a single point of failure, the central solution has
a computational complexity that scales cubically in the number of
vehicles and features. For this reason, here we propose a distributed solution
based on a combination of GMP and consensus algorithms.

\begin{figure*}[!t]
\centering \includegraphics[width=1\textwidth]{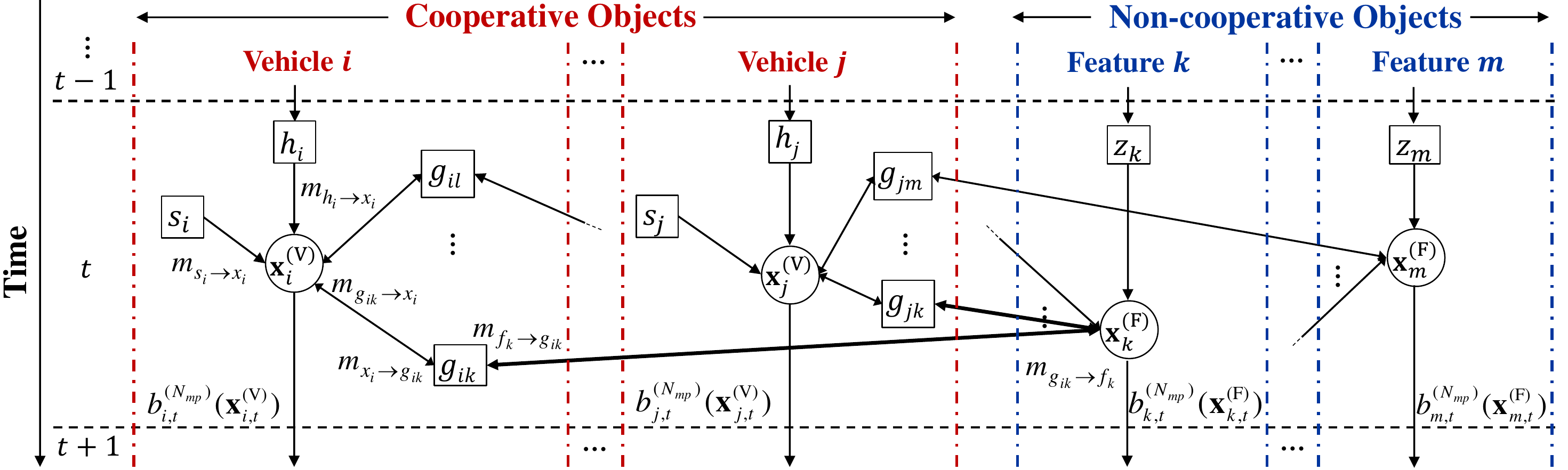}
\caption{FG of the joint posterior pdf (\ref{eq:cent_optimum_2}), showing the states of two vehicles $i,j \in \mathcal{V}$ and two features $k,m \in \mathcal{F}$. Vehicles (delimited by red dashed-dot lines) are active objects that cooperate through V2V links. Features (within blue dotted lines) are passive objects, not actively involved in the GMP, that are estimated through consensus by vehicles. As V2V cooperation is implicitly performed through jointly sensed features, in the FG the subgraphs associated to different vehicles are only connected by means of V2F measurements, e.g. by the thickest black arrows connecting vehicles $i$ and $j$ through feature $k$.}
\label{fig:fact_graph} 
\end{figure*} 
\begin{figure*}[b!]
\hrulefill\vspace*{-0.08cm}
\begin{equation}
%% The spacer can be tweaked to stop underfull vboxes.
\begin{split}
p(\boldsymbol{\theta}_{t}|\boldsymbol{\rho}_{1:t})\propto &\prod\limits _{i\in\mathcal{V}}\left[p(\boldsymbol{\rho}^{\mathrm{(GNSS)}}_{i,t}|\mathbf{x}^{(\mathrm{V})}_{i,t})\int p(\mathbf{x}^{(\mathrm{V})}_{i,t}|\mathbf{x}^{(\mathrm{V})}_{i,t-1})p(\mathbf{x}^{(\mathrm{V})}_{i,t-1}|\boldsymbol{\rho}_{1:t-1})d\mathbf{x}^{(\mathrm{V})}_{i,t-1}\prod\limits _{k\in\mathcal{F}_{i,t}}p(\boldsymbol{\rho}^{\mathrm{(V2F)}}_{i,k,t}|\mathbf{x}^{(\mathrm{F})}_{k,t},\mathbf{x}^{(\mathrm{V})}_{i,t})\right]\times \\
&\hspace{0.2cm}\prod\limits _{k\in\mathcal{F}}\int p(\mathbf{x}^{(\mathrm{F})}_{k,t}|\mathbf{x}^{(\mathrm{F})}_{k,t-1})p(\mathbf{x}^{(\mathrm{F})}_{k,t-1}|\boldsymbol{\rho}_{1:t-1})d\mathbf{x}^{(\mathrm{F})}_{k,t-1}\label{eq:cent_optimum_2}
\end{split}
\end{equation}\hrulefill
\end{figure*} %\FloatBarrier 
The distributed method enables the sequential evaluation, at each
vehicle $i\in\mathcal{V}$, of the marginal posterior pdfs $p(\mathbf{x}^{(\mathrm{V})}_{i,t}|\boldsymbol{\rho}_{1:t})$
and $p(\mathbf{x}^{(\mathrm{F})}_{k,t}|\boldsymbol{\rho}_{1:t})$, for all features
$k\in\mathcal{F}$. However, the GMP implementation is complicated
by the fact that features are passive objects and therefore they are not
actively involved in the estimation process. This means that each
vehicle $i$ has to calculate not only its own belief but also all
features' beliefs using only communication with neighboring vehicles.
To address this challenge, we propose a novel consensus-based GMP
method that enables the cooperation between vehicles for the distributed
evaluation of the all features' beliefs. 

An example of the proposed approach, and the related benefits, is in Fig. \ref{fig:ICP_example} for a scenario with one feature
jointly sensed by three vehicles. The figure shows the
localization accuracy drawn from the local beliefs when vehicles rely
only on their own GNSS (solid colored contours) and V2F measurements (dashed colored contours). On the other hand, in the ICP approach vehicles engage in a V2V cooperative localization of the feature and reach a consensus on the feature
location (black ellipse). This implicitly reflects on a significant improvement of vehicle position accuracies (colored ellipses). 

In the following, we discuss the distributed implementation of this method, by first describing the GMP solution to the specific estimation problem (Sec. \ref{subsec:GMP_alg}) and then the proposed
consensus-based approach (Sec. \ref{subsec:consensus_alg}). 

\subsection{Gaussian Message Passing Algorithm}\label{subsec:GMP_alg}

Taking into account the conditional independence of the measurements
and the static condition of the features, the posterior pdf (\ref{eq:cent_optimum})
can be factorized over vehicles and features as in (\ref{eq:cent_optimum_2})
at bottom of the next page. In order to derive the GMP, we first encode
\eqref{eq:cent_optimum_2} as a Factor Graph (FG) \cite{Loeliger2001}
and then derive the Sum-Product Algorithm (SPA) message passing rules. The FG of $p(\boldsymbol{\theta}_{t}|\boldsymbol{\rho}_{1:t})$
is depicted in Fig.~\ref{fig:fact_graph}, where the state of each
vehicle and feature is shown as a circle, while the factors in \eqref{eq:cent_optimum_2}
are shown as squares. For visualization purposes and to simplify the
notation, we introduce $h_{i}\triangleq p(\mathbf{x}^{(\mathrm{V})}_{i,t}|\mathbf{x}^{(\mathrm{V})}_{i,t-1})$
(i.e., vehicle state-transition pdf), $z_{k}\triangleq p(\mathbf{x}^{(\mathrm{F})}_{k,t}|\mathbf{x}^{(\mathrm{F})}_{k,t-1})$
(i.e., feature state-transition pdf), $s_{i}\triangleq p(\boldsymbol{\rho}^{\mathrm{(GNSS)}}_{i,t}|\mathbf{x}^{(\mathrm{V})}_{i,t})$
(i.e., the GNSS likelihood) and $g_{ik}\triangleq p(\boldsymbol{\rho}^{\mathrm{(V2F)}}_{i,k,t}|\mathbf{x}^{(\mathrm{F})}_{k,t},\mathbf{x}^{(\mathrm{V})}_{i,t})$ (i.e., the V2F measurement likelihood).

If the prior distributions of the vehicles and features are Gaussian,
if all measurements and state transition models are linear in the
state and have independent Gaussian noise, it can be shown that all
the messages in the FG are Gaussian \cite{loeliger2004}. Hence, the
SPA reverts to GMP, which has several benefits in terms of complexity
and convergence \cite{Van_Roy,weiss2001,weiss2001_2}. In \cite{weiss2001}, authors proved that if belief propagation converges in case of loopy graphs, then the posterior marginal belief mean vector converges to the optimal centralized estimate. Note that if the linearity and Gaussianity conditions are not fulfilled, particle-based approaches can be used, though these generally incur a significant computational and communication cost.

At each time $t$, the GMP scheme provides approximate marginal posteriors, which are represented by the beliefs of the vehicles' and features'
states, respectively $b_{i,t}(\mathbf{x}^{(\mathrm{V})}_{i,t})\approx p(\mathbf{x}^{(\mathrm{V})}_{i,t}|\boldsymbol{\rho}_{1:t})$
and $b_{k,t}(\mathbf{x}^{(\mathrm{F})}_{k,t})\approx p(\mathbf{x}^{(\mathrm{F})}_{k,t}|\boldsymbol{\rho}_{1:t})$.
Moreover, as the considered FG has cycles, the GMP algorithm becomes
iterative. Hence, the beliefs of the vehicle node $i\in\mathcal{V}$
and feature node $k\in\mathcal{F}$ at GMP iteration
$n=1,...,N_{\mathrm{mp}}$ are: 
\begin{align}
\begin{split}
b_{k,t}^{(n)}(\mathbf{x}^{(\mathrm{F})}_{k,t})\propto m_{z_{k}\to x_{k}}(\mathbf{x}^{(\mathrm{F})}_{k,t})\prod\limits _{i\in\mathcal{V}_{k,t}}m_{g_{ik}\to x_{k}}^{(n)}(\mathbf{x}^{(\mathrm{F})}_{k,t}),%\propto\mathcal{N}(\boldsymbol{\mu}_{f_{k,t}}^{(n)},\mathbf{C}_{f_{k,t}}^{(n)}),
\end{split}
\label{eq:beliefs_fea}\\
\begin{split}b_{i,t}^{(n)}(\mathbf{x}^{(\mathrm{V})}_{i,t})\propto\hspace{-0.05cm}m_{h_{i}\to x_{i}}(\mathbf{x}^{(\mathrm{V})}_{i,t})m_{s_{i}\to x_{i}}(\mathbf{x}^{(\mathrm{V})}_{i,t})\hspace{-0.15cm}\prod\limits _{k\in\mathcal{F}_{i,t}}\hspace{-0.2cm}m_{g_{ik}\to x_{i}}^{(n)}(\mathbf{x}^{(\mathrm{V})}_{i,t}),%\propto\mathcal{N}(\boldsymbol{\mu}_{x_{i,t}}^{(n)},\mathbf{C}_{x_{i,t}}^{(n)}),
\end{split}
\label{eq:beliefs_veh}
\end{align}
with $\mathcal{V}_{k,t}$ being the set of vehicles that acquire measurements of feature $k$ (assuming that the belief of feature $k\in\mathcal{F}$ in (\ref{eq:beliefs_fea}) 
is reset to a uniform distribution if the feature is not observed by any vehicle $i\in\mathcal{V}$). 

Note that the product of $L$ Gaussian pdfs over the same vector is also Gaussian (though not normalized) \cite{Loeliger2001}:
\begin{eqnarray}
\prod\limits _{\ell=1}^{L}\mathcal{N}(\boldsymbol{\mu}_{\ell},\mathbf{C}_{\ell})\propto\mathcal{N}(\tilde{\boldsymbol{\mu}},\mathbf{\tilde{C}}),\label{eq:gauss_prod}
\end{eqnarray}
with covariance $\tilde{\mathbf{C}}=(\sum_{\ell=1}^{L}\mathbf{C}_{\ell}^{-1})^{-1}$
and mean $\tilde{\boldsymbol{\mu}}=\tilde{\mathbf{C}}\cdot(\sum_{\ell=1}^{L}\mathbf{C}_{\ell}^{-1}\boldsymbol{\mu}_{\ell})$. This observation plays a key role for the evaluation of the beliefs in \eqref{eq:beliefs_fea}--\eqref{eq:beliefs_veh}, computed as follows. 
\begin{itemize}
\item \emph{Feature prediction message:} The predicted state of feature $k$ is represented by the message: 
\begin{align}
\begin{split} & m_{z_{k}\to x_{k}}(\mathbf{x}^{(\mathrm{F})}_{k,t})\hspace{-0.05cm}=\hspace{-0.1cm}\int\hspace{-0.1cm}p(\mathbf{x}^{(\mathrm{F})}_{k,t}|\mathbf{x}^{(\mathrm{F})}_{k,t-1}\hspace{-0.03cm})\ b_{k,t-1}^{(N_{\mathrm{mp}})}(\mathbf{x}^{(\mathrm{F})}_{k,t-1}\hspace{-0.03cm})\mathrm{d}\mathbf{x}^{(\mathrm{F})}_{k,t-1}\\
 & \hspace{0.1cm}=\mathcal{N}(\mathbf{A}\boldsymbol{\mu}_{x_{k,t-1}}^{(N_{\mathrm{mp}})},\mathbf{Q}^{(\mathrm{F})}_{k,t-1}+\mathbf{A}\mathbf{C}_{x_{k,t-1}}^{(N_{\mathrm{mp}})}\mathbf{A}^{\mathrm{T}}),
\end{split}
\label{eq:fea_predict_msg}
\end{align}
in which $\boldsymbol{\mu}_{x_{k,t-1}}^{(N_{\mathrm{mp}})}$ and $\mathbf{C}_{x_{k,t-1}}^{(N_{\mathrm{mp}})}$
are the mean and covariance of the feature belief at previous time instant, i.e., $b_{k,t-1}^{(N_{\mathrm{mp}})}(\mathbf{x}^{(\mathrm{F})}_{k,t-1}\hspace{-0.03cm})=\mathcal{N}(\boldsymbol{\mu}_{x_{k,t-1}}^{(N_{\mathrm{mp}})},\mathbf{C}_{x_{k,t-1}}^{(N_{\mathrm{mp}})})$
(similar notation will be used for other beliefs and messages).
\item \emph{Vehicle prediction message:} The predicted state of vehicle $i$ is represented by the message: 
\begin{align}
\begin{split} & m_{h_{i}\to x_{i}}(\mathbf{x}^{(\mathrm{V})}_{i,t})\hspace{-0.05cm}=\hspace{-0.1cm}\int\hspace{-0.1cm}p(\mathbf{x}^{(\mathrm{V})}_{i,t}|\mathbf{x}^{(\mathrm{V})}_{i,t-1}\hspace{-0.03cm})\ b_{i,t-1}^{(N_{\mathrm{mp}})}(\mathbf{x}^{(\mathrm{V})}_{i,t-1}\hspace{-0.03cm})\mathrm{d}\mathbf{x}^{(\mathrm{V})}_{i,t-1}\\
 & \hspace{0.1cm}=\mathcal{N}(\mathbf{B}\mathbf{a}_{i,t-1}+ \mathbf{A}\boldsymbol{\mu}_{x_{i,t-1}}^{(N_{\mathrm{mp}})},\mathbf{Q}^{(\mathrm{V})}_{i,t-1}+\mathbf{A}\mathbf{C}_{x_{i,t-1}}^{(N_{\mathrm{mp}})}\mathbf{A}^{\mathrm{T}}),
\end{split}
\label{eq:predict_msg}
\end{align}
in which $\boldsymbol{\mu}_{x_{i,t-1}}^{(N_{\mathrm{mp}})}$ and $\mathbf{C}_{x_{i,t-1}}^{(N_{\mathrm{mp}})}$
are the mean and covariance of the vehicle belief at previous time instant, i.e., $b_{i,t-1}^{(N_{\mathrm{mp}})}(\mathbf{x}^{(\mathrm{V})}_{i,t-1}\hspace{-0.03cm})=\mathcal{N}(\boldsymbol{\mu}_{x_{i,t-1}}^{(N_{\mathrm{mp}})},\mathbf{C}_{x_{i,t-1}}^{(N_{\mathrm{mp}})})$.
\item \emph{GNSS message:} The message $m_{s_{i}\to x_{i}}(\mathbf{x}^{(\mathrm{V})}_{i,t})$
is obtained according to the $i$th GNSS measurement
(\ref{eq:meas_GNSS_y_in}) and is a degenerate Gaussian with infinite variance in the velocity domain as the GNSS device provides only position information. Thereby, recalling that $\mathbf{p}^{(\mathrm{V})}_{i,t}=\mathbf{P}\mathbf{x}^{(\mathrm{V})}_{i,t}$ and $\mathbf{P}^{\dagger}=\mathbf{P}^{\mathrm{T}}$, the parameters $\boldsymbol{\mu}_{s_{i}\to x_{i}}$ and $\mathbf{C}_{s_{i}\to x_{i}}$ fulfill the following relations: $\mathbf{C}_{s_{i}\to x_{i}}^{-1}= \mathbf{P}^{\mathrm{T}}\mathbf{R}^{{\mathrm{(GNSS)}}^{-1}}_{i,t}\mathbf{P}$ and $\mathbf{C}_{s_{i}\to x_{i}}^{-1}\boldsymbol{\mu}_{s_{i}\to x_{i}}=\mathbf{C}_{s_{i}\to x_{i}}^{-1}\mathbf{P}^{\mathrm{T}}\boldsymbol{\rho}^{\mathrm{(GNSS)}}_{i,t}$. These parameters will be considered for the computation of the message below according to (\ref{eq:gauss_prod}). 
\item \emph{Message from vehicle $i$ to feature $k$:} The outgoing message at iteration $n$ from vehicle state $\mathbf{x}_{i}^{(\mathrm{V})}$ to factor $g_{ik}$ is obtained as:  
\begin{align}
\begin{split}m_{x_{i}\to g_{ik}}^{(n)}(\mathbf{x}^{(\mathrm{V})}_{i,t})\propto m_{h_{i}\to x_{i}}(\mathbf{x}^{(\mathrm{V})}_{i,t})m_{s_{i}\to x_{i}}(\mathbf{x}^{(\mathrm{V})}_{i,t})\\
\times\prod\limits _{\substack{m\in\mathcal{F}_{i,t}\setminus\{k\}}
}\hspace{-0.45cm}m_{g_{im}\to x{}_{i}}^{(n-1)}(\mathbf{x}^{(\mathrm{V})}_{i,t}),
\end{split}
\label{eq:out_msg_veh}
\end{align}
in which $m_{g_{im}\to x_{i}}^{(n-1)}(\mathbf{x}^{(\mathrm{V})}_{i,t})$ is the incoming message to vehicle computed as explained hereinafter. At BP iteration $n=1$, the outgoing message is set to $m_{x_{i}\to g_{ik}}^{(1)}(\mathbf{x}^{(\mathrm{V})}_{i,t})\propto m_{h_{i}\to x_{i}}(\mathbf{x}^{(\mathrm{V})}_{i,t})m_{s_{i}\to x_{i}}(\mathbf{x}^{(\mathrm{V})}_{i,t})$. 
The message is again Gaussian, i.e., $m_{x_{i}\to g_{ik}}^{(n)}(\mathbf{x}^{(\mathrm{V})}_{i,t})=\mathcal{N}(\boldsymbol{\mu}_{x_{i}\to g_{ik}}^{(n)},\mathbf{C}_{x_{i}\to g_{ik}}^{(n)})$, and it is then used to obtain the incoming message from factor $g_{ik}$ to feature state $\mathbf{x}^{(\mathrm{F})}_{k,t}$ as follows: 
\begin{equation} 
\hspace{-0.13cm} m_{g_{ik}\to x_{k}}^{(n)}\hspace{-0.04cm}(\hspace{-0.02cm}\mathbf{x}^{(\mathrm{F})}_{k,t}\hspace{-0.02cm})\hspace{-0.08cm}= \hspace{-0.17cm}\int\hspace{-0.17cm}p(\hspace{-0.02cm}\boldsymbol{\rho}^{\mathrm{(V2F)}}_{i,k,t}\hspace{-0.02cm}|\hspace{-0.02cm}\mathbf{x}^{(\mathrm{F})}_{k,t},\mathbf{x}^{(\mathrm{V})}_{i,t}\hspace{-0.02cm})\hspace{-0.01cm} m_{x_{i}\to g_{ik}}^{(n)}\hspace{-0.04cm}(\hspace{-0.02cm}\hspace{-0.02cm}\mathbf{x}^{(\mathrm{V})}_{i,t}\hspace{-0.02cm})\mathrm{d}\mathbf{x}^{(\mathrm{V})}_{i,t}.
\label{eq:incom_msg_fea}
\end{equation}
Note that the above message is a degenerate density with marginal pdf in the position domain Gaussian with mean $\boldsymbol{\rho}^{\mathrm{(V2F)}}_{i,k,t}+\mathbf{P}\boldsymbol{\mu}_{x_{i}\to g_{ik}}^{(n)}$ and covariance $\mathbf{R}^{\mathrm{(V2F)}}_{i,k,t}+\mathbf{P}\mathbf{C}_{x_{i}\to g_{ik}}^{(n)}\mathbf{P}^{\mathrm{T}}$. No information is available in the velocity domain (as variance over velocity tend to infinite). Thus, only the following parameters can be calculated: $(\mathbf{C}_{g_{ik}\to x_{k}}^{(n)})^{-1}\hspace{-0.08cm}= \mathbf{P}^{\mathrm{T}}(\mathbf{R}^{\mathrm{(V2F)}}_{i,k,t}+\mathbf{P}\mathbf{C}_{x_{i}\to g_{ik}}^{(n)}\mathbf{P}^{\mathrm{T}})^{-1}\mathbf{P}$ and $(\mathbf{C}_{g_{ik}\to x_{k}}^{(n)})^{-1}\boldsymbol{\mu}_{g_{ik}\to x_{k}}^{(n)}=(\mathbf{C}_{g_{ik}\to x_{k}}^{(n)})^{-1}\mathbf{P}^{\mathrm{T}}(\boldsymbol{\rho}^{\mathrm{(V2F)}}_{i,k,t}+\mathbf{P}\boldsymbol{\mu}_{x_{i}\to g_{ik}}^{(n)})$. These parameters are sufficient for the evaluation of the following message according to (\ref{eq:gauss_prod}). 

\item \emph{Message from feature $k$ to vehicle $i$:} The outgoing message from feature state $\mathbf{x}^{(\mathrm{F})}_{k,t}$ to factor $g_{ik}$ is:
\begin{align}
\begin{split}
m_{x_{k}\to g_{ik}}^{(n)}(\mathbf{x}^{(\mathrm{F})}_{k,t})\propto m_{z_{k}\to x_{k}}(\mathbf{x}^{(\mathrm{F})}_{k,t})\hspace{-0.05cm}\prod\limits _{\substack{l\in\mathcal{V}_{k,t}\setminus\{i\}}
}\hspace{-0.2cm}m_{g_{lk}\to x_{k}}^{(n)}(\mathbf{x}^{(\mathrm{F})}_{k,t}),
\end{split}
\label{eq:out_msg_fea}
\end{align}
in which $m_{x_{k}\to g_{ik}}^{(n)}(\mathbf{x}^{(\mathrm{F})}_{k,t})$$=$$\mathcal{N}(\boldsymbol{\mu}_{x_{k}\to g_{ik}}^{(n)},\mathbf{C}_{x_{k}\to g_{ik}}^{(n)})$. Note that
if vehicle $i$ is the only vehicle that observes feature $k$, the
message is equal to the belief computed at the previous time $t-1$. 
Now, the incoming message from factor $g_{ik}$ to vehicle state $\mathbf{x}_{i,t}^{(\mathrm{V})}$ is given by: 
\begin{equation}\label{eq:incom_msg_veh}
\hspace*{-0.2cm}m_{g_{ik}\to x_{i}}^{(n)}(\hspace*{-0.01cm}\mathbf{x}^{(\mathrm{V})}_{i,t}\hspace*{-0.01cm})\hspace*{-0.08cm}=\hspace*{-0.15cm}\int\hspace{-0.15cm}p(\boldsymbol{\rho}^{\mathrm{(V2F)}}_{i,k,t}\hspace*{-0.02cm}|\hspace*{-0.02cm}\mathbf{x}^{(\mathrm{F})}_{k,t},\mathbf{x}^{(\mathrm{V})}_{i,t}\hspace*{-0.01cm})\hspace*{-0.01cm} m_{x_{k}\to g_{ik}}^{(n)}(\hspace*{-0.01cm}\mathbf{x}^{(\mathrm{F})}_{k,t}\hspace*{-0.01cm})\mathrm{d}\mathbf{x}^{(\mathrm{F})}_{k,t}.
\end{equation}
Here, similarly to (\ref{eq:incom_msg_fea}) we get that: $(\mathbf{C}_{g_{ik}\to x_{i}}^{(n)})^{-1}= \mathbf{P}^{\mathrm{T}}(\mathbf{R}^{\mathrm{(V2F)}}_{i,k,t}+\mathbf{P}\mathbf{C}_{x_{k}\to g_{ik}}^{(n)}\mathbf{P}^{\mathrm{T}})^{-1}\mathbf{P}$ and $(\mathbf{C}_{g_{ik}\to x_{i}}^{(n)})^{-1}\boldsymbol{\mu}_{g_{ik}\to x_{i}}^{(n)}=(\mathbf{C}_{g_{ik}\to x_{i}}^{(n)})^{-1}\mathbf{P}^{\mathrm{T}}(-\boldsymbol{\rho}^{\mathrm{(V2F)}}_{i,k,t}+\mathbf{P}\boldsymbol{\mu}_{x_{k}\to g_{ik}}^{(n)})$.
\end{itemize}
Note that the beliefs of features and vehicles (\ref{eq:beliefs_fea})--(\ref{eq:beliefs_veh}),
as well as the outgoing messages (\ref{eq:out_msg_veh}) and (\ref{eq:out_msg_fea}), are all Gaussians with mean and covariance that can be evaluated from
the related incoming pdfs based on (\ref{eq:gauss_prod}).

Distributed implementation of the above GMP scheme requires all feature and vehicle nodes to make local computations, and exchange messages with neighbors. However, features are non-cooperative passive nodes that cannot make computations, neither can they communicate with vehicles. To enable fully distributed location estimation under these conditions, in the following we propose an average consensus algorithm \cite{Olfati-Saber-ProcIEEE2007} that is nested into the GMP in such a way to allow vehicles to evaluate the features' beliefs without their cooperation, by using only V2V broadcast communications.

\subsection{Consensus-based Evaluation of the Feature Beliefs and Outgoing Messages}\label{subsec:consensus_alg}

The product of measurement messages at feature $k$: 
\begin{equation}
u_{x_{k}}^{(n)}(\mathbf{x}^{(\mathrm{F})}_{k,t})\triangleq\hspace{-0.1cm}\prod\limits _{i\in\mathcal{V}_{k,t}}\hspace{-0.15cm}m_{g_{ik}\to x_{k}}^{(n)}(\mathbf{x}^{(\mathrm{F})}_{k,t}),\label{eq:umessage}
\end{equation}
is needed for the evaluation of the feature belief (\ref{eq:beliefs_fea}) and the outgoing message (\ref{eq:out_msg_fea}), which can be conveniently
rewritten as, respectively: 
\begin{align}
\begin{split}
b_{k,t}^{(n)}(\mathbf{x}^{(\mathrm{F})}_{k,t})\propto m_{z_{k}\to x_{k}}(\mathbf{x}^{(\mathrm{F})}_{k,t})\cdot u_{x_{k}}^{(n)}(\mathbf{x}^{(\mathrm{F})}_{k,t}),
\end{split}\label{eq:umessage-1}\\
\begin{split}
m_{x_{k}\to g_{ik}}^{(n)}(\mathbf{x}^{(\mathrm{F})}_{k,t})\propto m_{z_{k}\to x_{k}}(\mathbf{x}^{(\mathrm{F})}_{k,t})\cdot\frac{u_{x_{k}}^{(n)}(\mathbf{x}^{(\mathrm{F})}_{k,t})}{m_{g_{ik}\to x_{k}}^{(n)}(\mathbf{x}^{(\mathrm{F})}_{k,t})}.\end{split}\label{eq:out_cons_fea}
\end{align}
Unfortunately, as features are passive objects, they are not actively
involved in the GMP and they cannot merge the incoming messages into
\eqref{eq:umessage}. Messages can not even be calculated
by the vehicles individually; a cooperation between them is needed
using V2V communication links. Cooperation enables each vehicle to
evaluate all features' beliefs, even in case of no measurement between
vehicle and feature. 

Considering that all vehicles know the number of features in the network, the message $u_{x_{k}}^{(n)}(\mathbf{x}^{(\mathrm{F})}_{k,t})$ from \eqref{eq:umessage}
can be expressed as a product over all vehicles: 
\begin{equation}
u_{x_{k}}^{(n)}(\mathbf{x}^{(\mathrm{F})}_{k,t})=\hspace{-0.1cm}\prod\limits _{i\in\mathcal{V}}m_{g_{ik}\to x_{k}}^{(n)}(\mathbf{x}^{(\mathrm{F})}_{k,t}),\label{eq:umessage2}
\end{equation}
where each measurement message $m_{g_{ik}\to x_{k}}^{(n)}(\mathbf{x}^{(\mathrm{F})}_{k,t})$
is defined according to (\ref{eq:incom_msg_fea}) if $i\in\mathcal{V}_{k,t}$,
while for $i\notin\mathcal{V}_{k,t}$ it is set as a Gaussian pdf
with covariance matrix tending to infinity.

Now, since $u_{x_{k}}^{(n)}(\mathbf{x}^{(\mathrm{F})}_{k,t})\propto\mathcal{N}(\boldsymbol{\mu}_{u_{x_{k}}}^{(n)},\mathbf{C}_{u_{x_{k}}}^{(n)})$
according to (\ref{eq:gauss_prod}), both the mean $\boldsymbol{\mu}_{u_{x_{k}}}^{(n)}$
and the covariance $\mathbf{C}_{u_{x_{k}}}^{(n)}$ can be expressed
in terms of arithmetic average. Therefore, we propose to employ the
average consensus approach \cite{Olfati-Saber-ProcIEEE2007}, based
on successive refinements of local estimates at vehicles and information
exchange between neighbors, to cooperatively determine the first two
moments of $u_{x_{k}}^{(n)}(\mathbf{x}^{(\mathrm{F})}_{k,t})$, as detailed in the following.
\begin{itemize}
\item \emph{Computation of $\mathbf{C}_{u_{x_{k}}}^{(n)}$: } We introduce
a consensus variable $\boldsymbol{\Phi}_{i,x_{k}}^{(n,r)}$ for each
value of $i$, $k$ and $n$, that is initialized at consensus iteration
$r=0$ as:
\begin{align}
\boldsymbol{\Phi}_{i,x_{k}}^{(n,0)}=\left(\mathbf{C}_{g_{ik}\to x_{k}}^{(n)}\right)^{-1},
\end{align}
and subsequently updated according to the rule: 
\begin{equation}
\boldsymbol{\Phi}_{i,x_{k}}^{(n,r+1)}=\boldsymbol{\Phi}_{i,x_{k}}^{(n,r)}+\epsilon\sum\limits _{j\in\mathcal{J}_{i,t}}\left(\boldsymbol{\Phi}_{j,x_{k}}^{(n,r)}-\boldsymbol{\Phi}_{i,x_{k}}^{(n,r)}\right).\label{eq:cons_alg}
\end{equation}
We recall that $\mathcal{J}_{i,t}$ is the set of neighboring vehicles,
while the step-size $0<\epsilon<1/\Delta_{t}$ is chosen to ensure
convergence \cite{Olfati-Saber-ProcIEEE2007} to the average $1/N_{v}\sum_{i}\boldsymbol{\Phi}_{i,x_{k}}^{(n,0)}$.
Hence, after $N_{\mathrm{con}}$ consensus iterations we get: 
\begin{equation}
\boldsymbol{\Phi}_{i,x_{k}}^{(n,N_{\mathrm{con}})}\approx\frac{1}{N_{v}}\sum\limits _{i\in\mathcal{V}}\left(\mathbf{C}_{g_{ik}\to x_{k}}^{(n)}\right)^{-1},\label{eq:cons_convergence}
\end{equation}
from which we easily find that: 
\begin{align}
\begin{split} & \mathbf{C}_{u_{x_{k}}}^{(n)}\approx\left(N_{v}\boldsymbol{\Phi}_{i,x_{k}}^{(n,N_{\mathrm{con}})}\right)^{-1}.\end{split}
\label{eq:cons_mean_cov_fea1}
\end{align}

\item \emph{Computation of $\boldsymbol{\mu}_{u_{x_{k}}}^{(n)}$:} We again
introduce a consensus variable $\tilde{\boldsymbol{\Phi}}_{i,x_{k}}^{(n,r)}$
for each value of $i$, $k$ and $n$, initialized as: 
\begin{align}
\tilde{\boldsymbol{\Phi}}_{i,x_{k}}^{(n,0)}=\left(\mathbf{C}_{g_{ik}\to x_{k}}^{(n)}\right)^{-1}\boldsymbol{\mu}_{g_{ik}\to x_{k}}^{(n)},
\end{align}
and refined at iteration $r$ according to the same rule
as in \eqref{eq:cons_alg}. Using the same reasoning, we find that:
\begin{align}
\begin{split} & \boldsymbol{\mu}_{u_{x_{k}}}^{(n)}=N_{v}\mathbf{C}_{u_{x_{k}}}^{(n)}\tilde{\boldsymbol{\Phi}}_{i,x_{k}}^{(n,N_{\mathrm{con}})},%=\left(\sum\limits _{i\in\mathcal{V}}\left(\mathbf{C}_{g_{ik}\tox_{k}}^{(n)}\right)^{-1}\right)^{-1}.
\end{split}
\label{eq:cons_mean_cov_fea2}
\end{align}
in which $\mathbf{C}_{u_{x_{k}}}^{(n)}$ was obtained through \eqref{eq:cons_mean_cov_fea1}. 
\end{itemize}
Once an agreement is reached and \eqref{eq:cons_mean_cov_fea1}, \eqref{eq:cons_mean_cov_fea2}
are computed, each vehicle $i$ can evaluate the approximate marginal
posterior pdf of the feature $k$ at $n$th GMP iteration, $b_{k,t}^{(n)}(\mathbf{x}^{(\mathrm{F})}_{k,t})\propto\mathcal{N}(\boldsymbol{\mu}_{x_{k,t}}^{(n)},\mathbf{C}_{x_{k,t}}^{(n)})$,
from (\ref{eq:umessage-1}), with mean and covariance computed from
the moments of $u_{x_{k}}^{(n)}(\mathbf{x}^{(\mathrm{F})}_{k,t})$ and $m_{z_{k}\to x_{k}}(\mathbf{x}^{(\mathrm{F})}_{k,t})$
according to \eqref{eq:gauss_prod}. Next, the message $m_{x_{k}\to g_{ik}}^{(n)}(\mathbf{x}^{(\mathrm{F})}_{k,t})\propto\mathcal{N}(\boldsymbol{\mu}_{x_{k}\to g_{ik}}^{(n)},\mathbf{C}_{x_{k}\to g_{ik}}^{(n)})$
is obtained from \eqref{eq:out_cons_fea}, where the mean and the
covariance are given by: 
\begin{align}
\begin{split} & \boldsymbol{\mu}_{x_{k}\to g_{ik}}^{(n)}=\mathbf{C}_{x_{k}\to g_{ik}}^{(n)}\hspace{-0.1cm}\left(\hspace{-0.1cm}\left(\mathbf{C}_{x_{k,t}}^{(n)}\right)^{-1}\hspace{-0.2cm}\boldsymbol{\mu}_{x_{k,t}}^{(n)}\hspace{-0.1cm}-\hspace{-0.1cm}\left(\mathbf{C}_{g_{ik}\to x_{k}}^{(n)}\right)^{-1}\hspace{-0.2cm}\boldsymbol{\mu}_{g_{ik}\to x_{k}}^{(n)}\hspace{-0.1cm}\right),\\
 & \mathbf{C}_{x_{k}\to g_{ik}}^{(n)}=\left(\left(\mathbf{C}_{x_{k,t}}^{(n)}\right)^{-1}-\left(\mathbf{C}_{g_{ik}\to x_{k}}^{(n)}\right)^{-1}\right)^{-1}.
\end{split}
\label{eq:out_cons_fea_muC}
\end{align}
The proposed method is summarized in Algorithm \ref{alg_GMP}, where
the average consensus approach is nested into the GMP scheme discussed
in Sec. IV-A.

\renewcommand{\labelitemi}{$\textendash$}
\begin{algorithm}
\caption{Consensus-based Gaussian Message Passing}
\label{alg_GMP} 
\small
\begin{algorithmic}[1]
\State At time $t=0$ \textbf{Initialization}: 
\State \hspace{0.05cm} \textbf{vehicles} $i\in\mathcal{V}$ \textbf{in parallel} 
\State \hspace{0.05cm} initialize non-informative prior on vehicle {$p(\mathbf{x}^{(\mathrm{V})}_{i,0})$ }
\State \hspace{0.05cm} initialize non-informative prior on feature $p(\mathbf{x}^{(\mathrm{F})}_{k,0})$, $\forall k \in \mathcal{F}$
\State \hspace{0.05cm} \textbf{end parallel}
\For{$t=1\to T$} (time slot index)
\State \hspace{-0.5cm} \textbf{vehicles} $i\in\mathcal{V}$ \textbf{in parallel} 
\State \hspace{0.05cm} compute prediction messages $m_{h_i\to x_i}(\mathbf{x}^{(\mathrm{V})}_{i,t})$ according to
\Statex \hspace{0.62cm} the state-transition pdf as (\ref{eq:predict_msg})
\State \hspace{0.05cm} compute the message $m_{s_i\to x_i}(\mathbf{x}^{(\mathrm{V})}_{i,t})$ based on GNSS
\Statex \hspace{0.62cm} estimated position as in (\ref{eq:meas_GNSS_y_in}) 
\State \hspace{0.05cm} compute the initial outgoing message as
\Statex \hspace{0.65cm}\vspace{0.1cm}  $m_{x_{i}\to g_{ik}}^{(1)}(\mathbf{x}^{(\mathrm{V})}_{i,t})=\hspace{-0.05cm}m_{h_i\to x_i}(\mathbf{x}^{(\mathrm{V})}_{i,t})m_{s_i\to x_i}(\mathbf{x}^{(\mathrm{V})}_{i,t})$
\State \textbf{end parallel}
\For{$n=1\to N_{\mathrm{mp}}$} (GMP iteration index)
\State \hspace{-0.4cm}\textbf{vehicles} $i\in\mathcal{V}$ \textbf{in parallel}
\For{${k}\in\mathcal{F}$}	
		\State \hspace{-0.45cm} evaluate feature $k$ measurement message 
		\Statex \hspace{1.1cm} $m_{g_{ik}\to x_{k}}^{(n)}(\mathbf{x}^{(\mathrm{F})}_{k,t})$ according to (\ref{eq:incom_msg_fea})	
		\State \hspace{-0.45cm} compute the measurement message product $u_{x_{k}}^{(n)}(\mathbf{x}^{(\mathrm{F})}_{k,t})$
		\Statex \hspace{1.13cm} as (\ref{eq:cons_mean_cov_fea1}), (\ref{eq:cons_mean_cov_fea2}) by applying consensus algorithm (\ref{eq:cons_alg})
		\State \hspace{-0.45cm} update feature $k$ belief $b_{k,t}^{(n)}(\mathbf{x}^{(\mathrm{F})}_{k,t})$ as (\ref{eq:umessage-1}) 
		\State \hspace{-0.45cm} compute feature $k$ outgoing message
		\Statex \hspace{1.13cm} $m_{x_{k}\to g_{ik}}^{(n)}(\mathbf{x}^{(\mathrm{F})}_{k,t})$ as (\ref{eq:out_cons_fea}) by (\ref{eq:out_cons_fea_muC})
	\EndFor

\State \hspace{-0.45cm} compute vehicle $i$ incoming message $m_{g_{ik}\to x_{i}}^{(n)}(\mathbf{x}^{(\mathrm{V})}_{i,t})$ as (\ref{eq:incom_msg_veh})
\State \hspace{-0.45cm} update vehicle $i$  belief $b_{i,t}^{(n)}(\mathbf{x}^{(\mathrm{V})}_{i,t})$ as (\ref{eq:beliefs_veh})
\State \hspace{-0.45cm} compute vehicle $i$ outgoing message $m_{x_{i}\to g_{ik}}^{(n+1)}(\mathbf{x}^{(\mathrm{V})}_{i,t})$ as (\ref{eq:out_msg_veh}) 
\Statex \hspace{0.6cm} by applying (\ref{eq:gauss_prod})
\State \hspace{-0.45cm} \textbf{end parallel} 
\EndFor
\EndFor 
\end{algorithmic} 
\end{algorithm}

\section{ICP Performance Analysis}\label{sec:ICP Performance Analysis}

\subsection{Fundamental limits}

\label{subsec:centr_FIM}

In the following, we derive a lower bound to the cooperative localization
accuracy that is reached if and only if the $N_{v}$ vehicles and
the $N_{f}$ features are in full sensing and communication view (i.e.,
for all-to-all connectivity). To evaluate the vehicle positioning
accuracy, we first derive the overall Fisher Information Matrix (FIM),
$\mathbf{F}$, for the joint vehicle-feature state $\boldsymbol{\theta}_{t}$
and we then compute the submatrix of the FIM inverse, $\mathbf{C}_{t|t}=\mathbf{F}^{-1}$,
related to the single vehicle location. 

The total FIM can be obtained from (\ref{eq:Kalman_covariance}),
taking into account $\mathbf{H}_{t}$ and the block-diagonal
structure of $\mathbf{R}_{t}$. After some algebraic manipulations,
we get: 
\begin{equation}
\mathbf{F}=\begin{bmatrix}\mathbf{D} & \mathbf{E}\\
\mathbf{E}^{\mathrm{T}} & \mathbf{G}
\end{bmatrix},\label{eq:FIM}
\end{equation}
where the $4N_{v}\times4N_{v}$ matrix $\mathbf{D}=\text{blockdiag}(\mathbf{D}_{1},...,\mathbf{D}_{N_{v}})$
has submatrices $\mathbf{D}_{i}\in\mathbb{R}^{4\times4},$ $i\in\mathcal{V}$,
given by: 
\begin{equation}
\mathbf{D}_{i}=\mathbf{C}^{(\mathrm{V})^{-1}}_{i,t|t-1}+\mathbf{P}^{\mathrm{T}}\left(\mathbf{R}_{i,k,t}^{\mathrm{(GNSS)}^{-1}}+\sum\limits _{k\in\mathcal{F}_{i,t}}\mathbf{R}_{i,k,t}^{\mathrm{(V2F)}^{-1}}\right)\mathbf{P}.\label{eq: matrix Di}
\end{equation}
The $4N_{v}\times 4N_{f}$ matrix $\mathbf{E}=\left[\mathbf{E}_{ik}\right]$
is partitioned into blocks $\mathbf{E}_{ik}\in\mathbb{R}^{4\times4}$,
$i\in\mathcal{V},\ k\in\mathcal{F}$, defined as: 
\begin{equation}
\mathbf{E}_{ik}=\left\{ \begin{array}{lc}
-\mathbf{P}^{\mathrm{T}}\mathbf{R}_{i,k,t}^{\mathrm{(V2F)}^{-1}}\mathbf{P}, & \text{if }k\in\mathcal{F}_{i,t}\\
\mathbf{0}, & \textrm{otherwise}
\end{array}\right..\label{eq: matrix Eik}
\end{equation}

\noindent Moreover, the $4N_{f}\times4N_{f}$ matrix $\mathbf{G}=\text{blockdiag}(\mathbf{G}_{1},...,\mathbf{G}_{N_{f}})$
is built from the submatrices $\mathbf{G}_{k}\in\mathbb{R}^{4\times4},\forall k\in\mathcal{F}$,
such that: 
\begin{equation}
\mathbf{G}_{k}=\mathbf{C}^{(\mathrm{F})^{-1}}_{k,t|t-1}+\mathbf{P}^{\mathrm{T}}\sum\limits _{i\in\mathcal{V}_{k,t}}\mathbf{R}_{i,k,t}^{\mathrm{(V2F)}^{-1}}\mathbf{P}.\label{eq:FIM_last}
\end{equation}

To simplify the analysis, in the following we assume the prior covariance
matrices for vehicle $i$ and feature $k$ as, respectively, $\mathbf{C}^{\textsc{()}}_{i,t|t-1}=\mathrm{blockdiag}(\sigma_{p,\text{pr}}^{\textsc{(V)}^2}\mathbf{I}_{2},\sigma_{v,\text{pr}}^{\textsc{(V)}^2}\mathbf{I}_{2})$
and $\mathbf{C}^{(\mathrm{F})}_{k,t|t-1}=\mathrm{blockdiag}(\sigma_{p,\text{pr}}^{\textsc{(F)}^2}\mathbf{I}_{2},\sigma_{v,\text{pr}}^{\textsc{(F)}^2}\mathbf{I}_{2})$.
In addition, we assume the GNSS and V2F measurements as i.i.d. in
each subset, with covariance matrices $\mathbf{R}_{i,t}^{\mathrm{(GNSS)}}=\sigma_{\text{GNSS}}^{2}\mathbf{I}_{2}$
and $\mathbf{R}_{i,k,t}^{\mathrm{(V2F)}}=\sigma_{\mathrm{V2F}}^{2}\mathbf{I}_{2}$.
In this case, the FIM submatrices in (\ref{eq:FIM}) reduce to $\mathbf{D}=\mathbf{I}_{N_{v}}\otimes\mathrm{blockdiag}(\alpha_{p}^{\textsc{(V)}}\mathbf{I}_{2},\alpha_{v}^{\textsc{(V)}}\mathbf{I}_{2})$,
with $\alpha_{p}^{\textsc{(V)}}=N_{f}/\sigma_{\mathrm{V2F}}^{2}+1/\sigma_{p,\text{pr}}^{\textsc{(V)}^2}+1/\sigma_{\text{GNSS}}^{2}$ and $\alpha_{v}^{\textsc{(V)}}=1/\sigma_{v,\text{pr}}^{\textsc{(V)}^2}$,\vspace{0.05cm} $\mathbf{G}=\mathbf{I}_{N_{f}}\otimes\mathrm{blockdiag}(\alpha_{p}^{\textsc{(F)}}\mathbf{I}_{2},\alpha_{v}^{\textsc{(F)}}\mathbf{I}_{2})$,
with $\alpha_{p}^{\textsc{(F)}}=N_{v}/\sigma_{\mathrm{V2F}}^{2}+1/\sigma_{p,\text{pr}}^{\textsc{(F)}^2}$ and $\alpha_{v}^{\textsc{(F)}}=1/\sigma_{v,\text{pr}}^{\textsc{(F)}^2}$,
and $\mathbf{E}=-1/\sigma_{\mathrm{V2F}}^{2}\mathbf{1}_{N_{v}\times N_{f}}\otimes\mathbf{P}^{\mathrm{T}}\mathbf{P}$,
where $\mathbf{1}_{N_{v}\times N_{f}}$ is an $N_{v}\times N_{f}$
matrix of all ones. 

The equivalent FIM (EFIM) for the vehicles' states is given by the
Schur complement \cite{horn1987matrix}: 
\begin{equation}
\begin{split}\mathbf{F}^{(\mathrm{V})} & =\hspace{-0.07cm}\mathbf{D}-\mathbf{E}\mathbf{G}^{-1}\mathbf{E}^{\mathrm{T}}\\
 & =\hspace{-0.07cm}\mathbf{D}-\beta(\mathbf{1}_{N_{v}\times N_{f}}\otimes\mathbf{P}^{\mathrm{T}}\mathbf{P})(\mathbf{1}_{N_{v}\times N_{f}}\otimes\mathbf{P}^{\mathrm{T}}\mathbf{P})^{\mathrm{T}}\\
 & =\mathbf{I}_{N_{v}}\hspace{-0.1cm}\otimes\hspace{-0.1cm}\begin{bmatrix}\alpha_{p}^{\textsc{(V)}}\mathbf{I}_{2} & \hspace{-0.1cm}\mathbf{0}_{2\times2}\\
\mathbf{0}_{2\times2} & \hspace{-0.1cm}\alpha_{v}^{\textsc{(V)}}\mathbf{I}_{2}
\end{bmatrix}\hspace{-0.1cm}-\hspace{-0.1cm}\tilde{\beta}\mathbf{1}_{N_{v}\times N_{v}}\hspace{-0.1cm}\otimes\mathbf{P}^{\mathrm{T}}\mathbf{P}
\end{split}
\label{eq:EFIM}
\end{equation}
with $\beta=1/(\alpha_{p}^{\textsc{(F)}}\sigma_{\mathrm{V2F}}^{4})$,
$\tilde{\beta}=N_{f}\beta$ and where we made use of $(\mathbf{1}_{N_{v}\times N_{f}}\otimes\mathbf{P}^{\mathrm{T}}\mathbf{P})(\mathbf{I}_{N_{f}}\otimes\mathrm{blockdiag}(1/\alpha_{p}^{\textsc{(F)}}\mathbf{I}_{2},1/\alpha_{v}^{\textsc{(F)}}\mathbf{I}_{2}))=1/\alpha_{p}^{\textsc{(F)}}\left(\mathbf{1}_{N_{v}\times N_{v}}\otimes\mathbf{P}^{\mathrm{T}}\mathbf{P}\right)$ and $(\mathbf{1}_{N_{v}\times N_{f}}\otimes\mathbf{P}^{\mathrm{T}}\mathbf{P})(\mathbf{1}_{N_{v}\times N_{f}}\otimes\mathbf{P}^{\mathrm{T}}\mathbf{P})^{\mathrm{T}}=N_{f}\left(\mathbf{1}_{N_{v}\times N_{v}}\otimes\mathbf{P}^{\mathrm{T}}\mathbf{P}\right)$.
The inverse of the EFIM, $\mathbf{C}^{(\mathrm{V})}_{t|t}=\mathbf{F}^{(\mathrm{V})^{-1}}$,
represents the lower bound on the mean square error (MSE) matrix of
the vehicles' position estimates and it is of the form: 
\begin{equation}
\mathbf{C}^{(\mathrm{V})}_{t|t}\hspace{-0.07cm}=\hspace{-0.07cm}\mathbf{I}_{N_{v}}\hspace{-0.05cm}\otimes\hspace{-0.05cm}\begin{bmatrix}\alpha_{p}^{\textsc{(V)}^{-1}}\mathbf{I}_{2}\hspace{-0.1cm} & \mathbf{0}_{2\times2}\\
\mathbf{0}_{2\times2}\hspace{-0.1cm} & \alpha_{v}^{\textsc{(V)}^{-1}}\mathbf{I}_{2}
\end{bmatrix}\hspace{-0.05cm}+\eta\mathbf{1}_{N_{v}\times N_{v}}\hspace{-0.1cm}\otimes\mathbf{P}^{\mathrm{T}}\mathbf{P},
\end{equation}
where a simple association yields to $\eta=(1/\alpha_{p}^{\textsc{(V)}})/(\alpha_{p}^{\textsc{(V)}}/\tilde{\beta}-N_{v})$.
Hence, the posterior covariance matrix of the position-velocity estimate
for any vehicle $i$ is: 
\begin{equation}
\mathbf{C}^{(\mathrm{V})}_{t|t}=\begin{bmatrix}\sigma_{p,\text{post}}^{{\textsc{(V)}}^2}\mathbf{I}_{2} & \mathbf{0}_{2\times2}\\
\mathbf{0}_{2\times2} & \sigma_{v,\text{post}}^{{\textsc{(V)}}^2}\mathbf{I}_{2}
\end{bmatrix}=\begin{bmatrix}(\alpha_{p}^{\textsc{(V)}^{-1}}+\eta)\mathbf{I}_{2} & \mathbf{0}_{2\times2}\\
\mathbf{0}_{2\times2} & \alpha_{v}^{\textsc{(V)}^{-1}}\mathbf{I}_{2}
\end{bmatrix},
\end{equation}
where $\alpha_{p}^{\textsc{(V)}^{-1}}$ is the expected uncertainty if the features
behaved as anchors, i.e., their locations were perfectly known and
thus $\sigma_{p,\text{pr}}^{\textsc{(F)}}=0$.

Focusing on vehicle position only, after some manipulations we get:
\begin{equation}\label{eq:veh_sigma_final}
\sigma_{p,\text{post}}^{{\textsc{(V)}}^2}=\frac{1}{\alpha_{p}^{\textsc{(V)}}}\left(\hspace{-0.1cm}1\hspace{-0.02cm}+\hspace{-0.02cm}\frac{N_{f}/\sigma_{\mathrm{V2F}}^{2}}{N_{v}/\sigma_{p,\text{pr}}^{\textsc{(V)}^2}\hspace{-0.02cm}+\hspace{-0.02cm}N_{v}/\sigma_{\text{GNSS}}^{2}\hspace{-0.02cm}+\hspace{-0.02cm}\alpha_{p}^{\textsc{(V)}}\sigma_{\mathrm{V2F}}^{2}/\sigma_{p,\text{pr}}^{\textsc{(F)}^2}}\right).
\end{equation}

\subsection{Performance Scaling}

\author{\label{subsec:centr_scaling}}

Based on the result in \eqref{eq:veh_sigma_final}, we consider the
following limiting cases (see Appendix for derivation): 

\textit{Large number of vehicles, $N_{v}\rightarrow\infty$}: 
\begin{equation}
\sigma_{p,\text{post}}^{{\textsc{(V)}}^2}\rightarrow\frac{1}{\alpha_{p}^{\textsc{(V)}}}=\frac{1}{N_{f}/\sigma_{\mathrm{V2F}}^{2}+1/\sigma_{p,\text{pr}}^{\textsc{(V)}^2}+1/\sigma_{\text{GNSS}}^{2}},\label{eq:high_Nv}
\end{equation}
which is the accuracy reached when all features act as anchors. Since
each feature is observed by an infinite number of vehicles, its location
becomes perfectly known.

\textit{Large number of features,} $N_{f}\rightarrow\infty$: 
\begin{equation}
\sigma_{p,\text{post}}^{{\textsc{(V)}}^2}\rightarrow0.\label{eq:high_Nf}
\end{equation}
In this case all vehicles' locations become certain, as features behave
like anchors, even if still 'virtual' (i.e., possibly uncertain to
some extent), provided that there are many of them.

\textit{Small V2F measurement variance,} $\sigma_{\mathrm{V2F}}^{2}\rightarrow0$:
\begin{equation}
\sigma_{p,\text{post}}^{{\textsc{(V)}}^2}\rightarrow\frac{1}{N_{v}/\sigma_{p,\text{pr}}^{\textsc{(V)}^2}+N_{v}/\sigma_{\text{GNSS}}^{2}+N_{f}/\sigma_{p,\text{pr}}^{\textsc{(F)}^2}},\label{eq:low_sigma_r}
\end{equation}
so the performance reaches a limiting value. When also the features'
locations are perfectly known, i.e., $\sigma_{p,\text{pr}}^{\textsc{(F)}^2}\rightarrow0$,
we get $\sigma_{p,\text{post}}^{{\textsc{(V)}}^2}\rightarrow0$. It follows that
good measurements \textit{and} good prior feature information are
required to have good positioning, when there are not many features,
as intuitively expected.

\textit{Small feature prior uncertainty,} $\sigma_{p,\text{pr}}^{\textsc{(F)}^2}\rightarrow0$:
\begin{equation}
\sigma_{p,\text{post}}^{{\textsc{(V)}}^2}\rightarrow\frac{1}{\alpha_{p}^{\textsc{(V)}}},\label{eq:low_sigma_f}
\end{equation}
which means that features are like true anchors. 
\begin{enumerate}
\item \textit{Small vehicle prior uncertainty}, $\sigma_{p,\text{pr}}^{\textsc{(V)}^2}\rightarrow0$,
or \textit{GNSS position uncertainty}, $\sigma_{\text{GNSS}_{}}^{2}\rightarrow0$:
\begin{equation}
\sigma_{p,\text{post}}^{{\textsc{(V)}}^2}\rightarrow0.\label{eq:low_sigma_v_or_GNSS}
\end{equation}

\noindent In this case cooperation is not worth, as stand-alone positioning at each vehicle is enough accurate. 

\end{enumerate}

\section{Implementation Aspects}
In this section, we comment on implementation aspects related to data
association, V2V communication, complexity and measurement synchronization.

\subsection{Cooperative Data Association}\label{sec:Cooperative Data Assoc.} 
Some form of data association
is required for the implementation of the proposed cooperative localization
approach. In particular, each vehicle needs to track features in view,
and associate measurements to features. Several approaches are available in the multi-target tracking literature \cite{BarShalom90}, accounting for the arrival of new features and the removal of features no longer in view. In addition to per-vehicle data association, vehicles must agree on
a common set of features.  Approaches for this cooperative
data association problem exist as well \cite{chen2005distributed,MeyerBWH16,wang2016distributed}.
Some of these approaches are based on FGs, and can thus be
incorporated in the proposed localization algorithm. %In contrast to the above mentioned works, in our setting both sensors (vehicles) and targets (features) are mobile, both with uncertain positions. 
In our work, we do not explicitly treat the data association problem, but rather assume that the local sensors can provide unique semantic labels for each detected feature (e.g., for a camera sensor, this label could be of the form ``person with green jacket and blue trousers"), based on which the proposed positioning algorithm can be performed. In that sense, our algorithm provides a lower bound on the location error for a more practical algorithm with data association. 
This creates several challenges that must be addressed, but are outside
the scope of this paper.

\subsection{Communication Overhead}

\label{sec:Comm. Overhead} Irrespective of the form of data association,
the proposed localization method requires significant communication
between vehicles, as discussed below.
\begin{enumerate}
\item \emph{Cooperative data association:} During this phase, vehicles
decide on which local feature identifier corresponds to local feature
identifiers of other vehicles. Each vehicle can thus maintain a list
of vehicles for each feature and a list of features that it shares
with other vehicles. Such lists remove the need for all vehicles to
keep track of all features. Consensus-based methods can be applied \cite{wang2016distributed}. %Assuming vehicles that can observe the
%same feature are always in communication range, the cooperative data association requires one broadcast per vehicle. 
\item \emph{BP iterations:} Once each vehicle has knowledge of features
and associated vehicles that agree on the same features, the BP iterations
commence. Each BP iteration, as shown in Algorithm \ref{alg_GMP},
mainly consists of consensus iterations. During each consensus iteration,
each vehicle broadcasts feature-related information. Each broadcast
would comprise transmitter ID, transmitter belief, feature identifier
per feature, feature belief per feature. 
\end{enumerate}
From the above discussion, it is clear that the total number of broadcasts
per vehicle is dominated by the consensus, and thus scales as $\mathcal{O}(N_{\mathrm{mp}}N_{\mathrm{con}})$,
where $N_{\mathrm{mp}}$ denotes the number of BP iterations and $N_{\mathrm{con}}$
the number of consensus iterations per BP iteration. Considering a
data rate of $R$ bits/s, the time required for communication is lower
bounded by: 
\begin{equation}
T_{s}\ge N_{\mathrm{mp}}N_{\mathrm{con}}N_{f}N_{\mathrm{nei}}N_{b}/R,\label{eq:comm_ov}
\end{equation}
where $N_{\mathrm{nei}}$ is the number of neighboring vehicles and
$N_{b}$ is the number of bits needed to describe the belief of a
feature. As an example, in the case of using the IEEE 802.11p V2V
standard, with $R=6$ Mbit/s, $N_{\mathrm{nei}}=10$ neighbors, $N_{f}=20$
features, $N_{b}=100$ bits, and $T_{s}=1$ s, we find that $N_{\mathrm{mp}}N_{\mathrm{con}}\le300$,
which is a reasonable number, as we will see during the performance
evaluation.

\subsubsection*{Remark}

To reduce the communication overhead and delay, the value of $N_{\mathrm{mp}}$
can be made adaptive. In our case, at vehicle $i$, we stop the GMP
iterations when $ ||\boldsymbol{\mu}^{(n+1)}_{x_{i,t}} -\boldsymbol{\mu}^{(n)}_{x_{i,t}}||<\gamma_{\mathrm{mp}}$ and $||\mathbf{C}^{(n+1)}_{x_{i,t}} -\mathbf{C}^{(n)}_{x_{i,t}}||^{1/2}<\gamma_{\mathrm{mp}}$, $\forall i\in \mathcal{V}$, with $\gamma_{\mathrm{mp}}$ being a threshold and $\boldsymbol{\mu}^{(n)}_{x_{i,t}}$ and $\mathbf{C}^{(n)}_{x_{i,t}}$ being respectively the mean and the covariance of the $i$th vehicle belief $b_{i,t}^{(n)}(\mathbf{x}^{(\mathrm{V})}_{i,t})$. Similarly, the value of $N_{\mathrm{con}}$ can be made adaptive. In our case $||\tilde{\boldsymbol{\Phi}}^{(n,r+1)}_{i,x_{k}} -\tilde{\boldsymbol{\Phi}}^{(n,r)}_{i,x_{k}}|| <\gamma_{\mathrm{con}}$ and $ ||\boldsymbol{\Phi}^{(n,r+1)}_{i,x_{k}} -\boldsymbol{\Phi}^{(n,r)}_{i,x_{k}}||^{1/2} <\gamma_{\mathrm{con}}$, $\forall i\in \mathcal{V}$ and $\forall k\in \mathcal{F}$, with a threshold $\gamma_{\mathrm{con}}$ and $\tilde{\boldsymbol{\Phi}}^{(n,r)}_{i,x_{k}}$ and $\boldsymbol{\Phi}^{(n,r)}_{i,x_{k}}$ being respectively the variables used to determine the first two moments of $u_{x_k}^{(n)}(\mathbf{x}^{(\mathrm{F})}_{k})$ (\ref{eq:umessage2}), needed for the evaluation of the $k$th feature belief $b_{k,t}^{(n)}(\mathbf{x}^{(\mathrm{F})}_{k})$. 

\subsection{Computational Complexity}

\label{sec:Complexity} Ignoring the complexity of the data association,
the per-vehicle computational complexity of the proposed method is
relatively modest in comparison with the centralized approach from
Section \ref{sec:Centr_solution}. In particular, the consensus iterations
require only additions of vectors, which scales linearly in the number
of features. In addition each vehicle must invert $N_{f}+1$ covariance
matrices of dimension $2\times2$, so that the total complexity per
time slot scales as $\mathcal{O}(N_{v}N_{\mathrm{mp}}(N_{\mathrm{con}}N_{f}+8N_{f}))$.
In contrast, the complexity of the centralized approach is dominated
by the inversion of covariance matrices, with a total complexity per
time slot scaling as $\mathcal{O}((N_{v}+N_{f})^{3})$.

\begin{figure*}[!t]
\centering \includegraphics[width=0.99\textwidth]{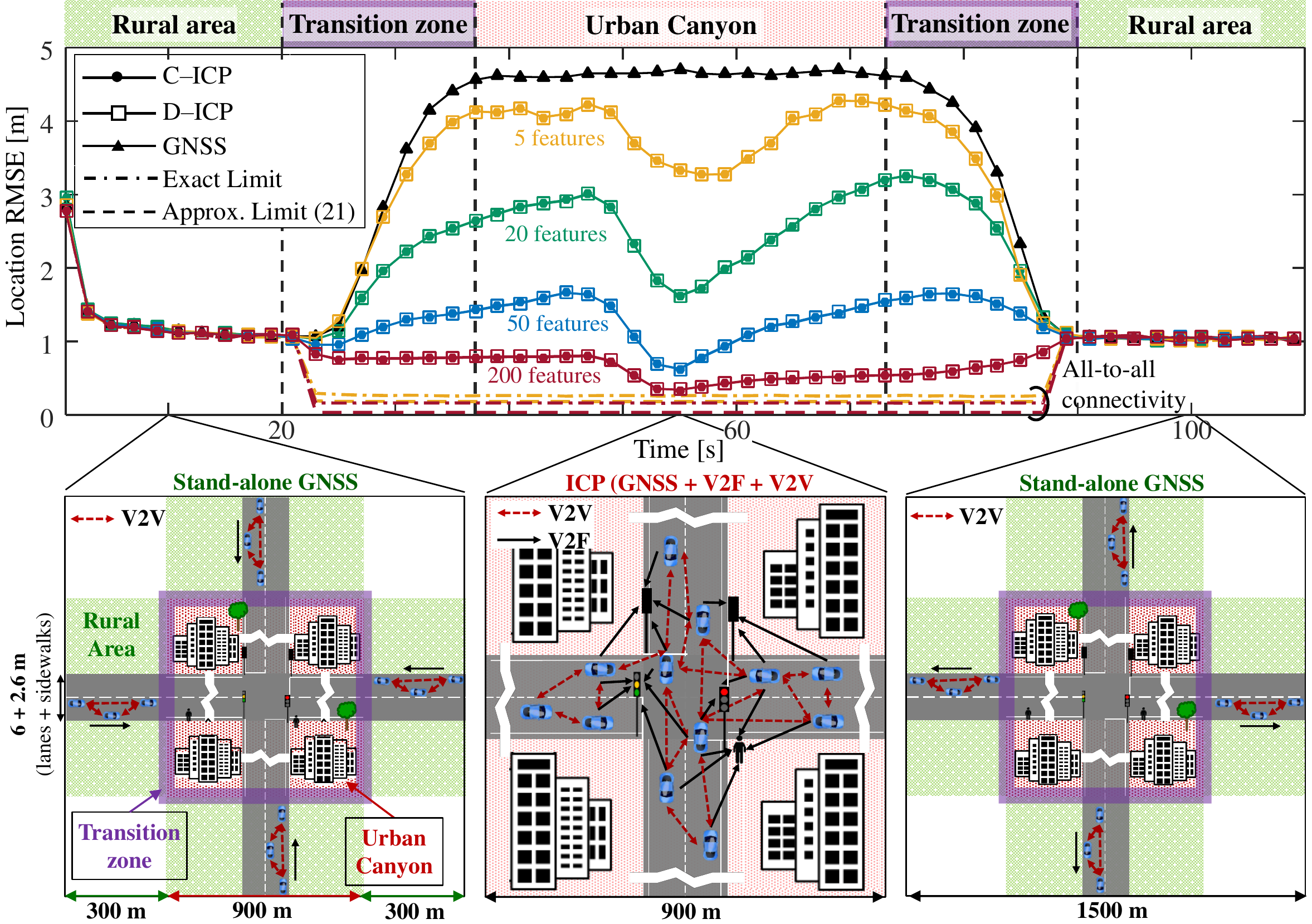}
\caption{ICP performance in a crossroad scenario with $N_v=12$ vehicles driving through a $1.5$ km $\times 1.5$ km area. The scenario is pictured in the bottom figures at time instant $t=10$ s (vehicles are driving through the rural area towards the crossroad), $t=55$ s (vehicles are crossing the urban canyon) and $t=100$ s (vehicles are back in the rural area). In the rural area only GNSS is available, while in the urban canyon $N_f \in \{5, 20, 50, 200\}$ features are jointly sensed by the ICP-enabled vehicles to augment the GNSS performance. The violet box highlights the transition from one area to the other. In the top figure, the ICP accuracy versus time is compared with stand-alone GNSS and with the lower bound for all-to-all V2V/V2F connectivity.}
\label{fig:results_fig4} 
\end{figure*}

\subsection{Measurement Synchronization}\label{sec:Meas Synchro.} Ideally, observations with respect to sensed features (e.g., relative positions derived out of range and azimuth angle measurements) should be isochronous and spatially coherent for a common time $t$ before performing consensus iterations. In particular, one mostly has to guarantee that the measurements associated with a group of cooperating vehicles fall in a sufficiently short period of time, which should be reasonably small in comparison with the positioning sampling time $T_s$. Considering that both the refresh period of perceptual sensors such as RADARs or LIDARs (typically, a few tens of ms)~\cite{DPMuller_2017} and the nominal broadcast period of awareness messages (typically, on the order of 100 ms for IEEE 802.11p/ITS-G5 in the steady-state regime or even below in case of event-triggered transmissions) are lower than $T_s$ (typically, on the order of 1 s), the assumption of quasi-isochronous measurements reasonably holds. Particularly, a simple criterion is detecting if new measurement data is outdated for integration in the fusion process and thus it is not sufficiently aligned in time with the measurements of cooperating vehicles. Other schemes \cite{Denis2015} are feasible to investigate, but do not fall in scope of this paper.

\section{Performance Evaluation}\label{sec:Performance Evaluation}

In this section the ICP performance is assessed in two different scenarios. A cross-road area is first simulated in Sec. \ref{sec:Crossroad_Performance Evaluation} with static features and heterogeneous positioning conditions in terms of feature density and V2V connectivity. This scenario is used to investigate the ICP accuracy for varying number of features and vehicles, and also to validate the analytical bound derived in Sec. \ref{sec:ICP Performance Analysis}. A more complex scenario is then introduced in Sec. \ref{sec:SUMO_Performance Evaluation}, where vehicular and pedestrian traffic is simulated over a real urban map using the SUMO simulator. This use-case is considered to validate the ICP method in more realistic traffic conditions, with mobile features and vehicles using different types of GNSS device with  significant diversity of location accuracy.

\subsection{Simulated crossroad scenario with mixed rural/urban areas}\label{sec:Crossroad_Performance Evaluation} 

\textbf{\textit{Settings.}} We first consider the crossroad scenario in Fig.~\ref{fig:results_fig4} bottom-left map, where the total length of each road is $1.5$ km and the center of the intersection is at position $\mathbf{c} = [750\text{ m},750\text{ m}]$. Lane and sidewalk widths are respectively set to $3$ m and $1.3$ m. As illustrated in the three different time frames at the bottom of  Fig.~\ref{fig:results_fig4}, the scenario involves $N_v$ vehicles, grouped in four clusters of $N_v/4$ vehicles each, that enter at time $t=0$ from the four corners of the area, drive straight ahead along their respective lanes and exit on the opposite sides, crossing in the middle. Each vehicle drives through three different areas: a rural area (first road section of $300$ m), urban canyon (the central section of $900$ m) and again a rural area (last $300$ m). Since vehicles need some time to enter/leave different areas, there is a transitory interval in which different vehicles are in different areas, with duration that depends on the specific parameter settings. 

In terms of vehicle dynamics, for each vehicle we set the initial velocity to $\mathbf{v}^{(\mathrm{V})}_{0}=\textbf{0}$ km/h. The mean acceleration $\mathbf{a}_{i,t}$ in (\ref{eq:veh_motion_model}) is initialized to $1.4$ m/s$^2$ in the driving direction at $t=0$ and kept constant until the vehicle reaches a velocity of 50 km/h, then it is set to 0 m/s$^2$ (i.e., the average driving velocity is 50 km/h). Since vehicles move along roads, the acceleration uncertainty in the direction of road, $\sigma_{a_{i,||}} = 0.3$ m/s$^2$, is assumed to be greater than the one in the orthogonal direction, $\sigma_{a_{i,\perp}} = 10^{-4}$ m/s$^2$. Thus, depending on the driving direction of the vehicle, the acceleration uncertainties along $x$ and $y$ axes, respectively $\sigma_{a_{xi}}$ and $\sigma_{a_{yi}}$, are defined. The sampling time is $T_s = 1$ s. 
The process noise covariance matrix from \eqref{eq:veh_motion_model} is set as $\mathbf{Q}_{i,t-1}^{(\mathrm{V})}=\mathbf{B}\tilde{\mathbf{Q}}^{(\mathrm{V})}_{i,t-1}\mathbf{B}^T$, with:
\begin{equation}
\tilde{\mathbf{Q}}_{i,t-1}^{(\mathrm{V})}=
\begin{bmatrix}
\sigma_{a_{i,||}}^2 & 0\\
0 & \sigma_{a_{i,\perp}}^2
\end{bmatrix}.
\end{equation}

For positioning, in the rural area vehicles rely solely on GNSS, while in the urban canyon they can also use features, which are randomly deployed over the area. In this scenario, features are assumed to be static, thus their mobility model in (\ref{eq:fea_motion_model}) reduces to $\mathbf{x}_{k,t}^{(\mathrm{F})}= \mathbf{A}\mathbf{x}_{k,t-1}^{(\mathrm{F})}=\mathbf{x}_{k,t-1}^{(\mathrm{F})}$ as $\mathbf{v}_{k,t}^{(\mathrm{F})}= \mathbf{v}_{k,t-1}^{(\mathrm{F})}=\textbf{0}$ km/h. Note that in all the simulated methods, vehicle dynamics are incorporated using either a Kalman filter or the GMP. 

The GNSS measurement covariance matrix at each vehicle is $\mathbf{R}^{(\mathrm{GNSS})}_{i,t} =\sigma_{\text{GNSS}}^2\mathbf{I}_2$, with $\sigma_{\text{GNSS}}=2$ m in the rural area and $\sigma_{\text{GNSS}}=15$ m in the urban canyon. The V2F measurement covariance matrix is $\mathbf{R}^{(\mathrm{V2F})}_{i,k,t} = \sigma_{\mathrm{V2F}}^2\mathbf{I}_2$, with $\sigma_{\mathrm{V2F}} = 0.5$ m. Finally, the communication and sensing ranges at each vehicle are set to $R_c = 150$ m and $R_s = 50$ m, respectively. The consensus step-size parameter is set to $\epsilon = 0.99/\Delta_t$, while the threshold on the GMP and consensus convergence are set to $\gamma_{\mathrm{mp}}=\gamma_{\mathrm{con}}=10^{-2}$.

In the following, the positioning performance is evaluated through Monte Carlo simulations, in terms of (i) the root mean square error (RMSE) of the position estimate and (ii) the delay of the fix delivery (measured in terms of the number of GMP iterations $N_{\mathrm{mp}}$ and consensus iterations $N_{\mathrm{con}}$). Three methods are compared, namely the stand-alone GNSS, the centralized and distributed versions of the proposed ICP method. 

\textbf{\textit{Numerical results.}} We first investigate the performance for a fixed number of vehicles $N_v=12$ and a varying number of features. Fig.~\ref{fig:results_fig4} shows the position RMSE of the vehicles as a function of time, for the three positioning methods. Snapshots of the V2V/V2F connectivity are shown at the bottom for time instants $t=10$ s (when vehicles are driving through the rural area towards the crossroad), $t=55$ s (in the urban canyon) and $t=100$ s (back to the rural area). Note that the exponential decay of the RMSE in the first few seconds of simulation results is due to transient effects. When vehicles use stand-alone GNSS, a severe performance degradation is observed as soon as vehicles enter the transition zone. The proposed algorithm can counter this degradation, especially when many features are available. The centralized and distributed ICP methods, namely C-ICP and D-ICP, lead to nearly identical performance, indicating that the proposed solution does not suffer from cycles in the FG. Moreover, assuming all-to-all V2V and V2F connectivity, the exact lower bound (dashed-dot line), obtained from (\ref{eq:Kalman_covariance}) for $\mathcal{F}_{i,t}=\mathcal{F}$ and $\mathcal{V}_{k,t}=\mathcal{V},\ \forall t$, and the approximated one (dashed line), from (\ref{eq:veh_sigma_final}), are evaluated for $N_f \in \{5, 200\}$. For the latter limit, variances are computed by approximating the prior/measurement covariances as diagonal matrices with entries determined as sample averages over the two spatial dimensions (for both vehicles and features). It can be seen that when the connectivity is high, a moderate number of features and vehicles (respectively, $5$ and $12$) is enough to obtain a centimeter-level accuracy. As predicted from the theoretical analysis in Sec. \ref{sec:Centr_solution}, if the number of features is high, e.g., $N_f=200$, the vehicle location accuracy tends to zero (see (\ref{eq:high_Nf})).

\begin{figure}[!t]
\centering 
\includegraphics[width=1\columnwidth]{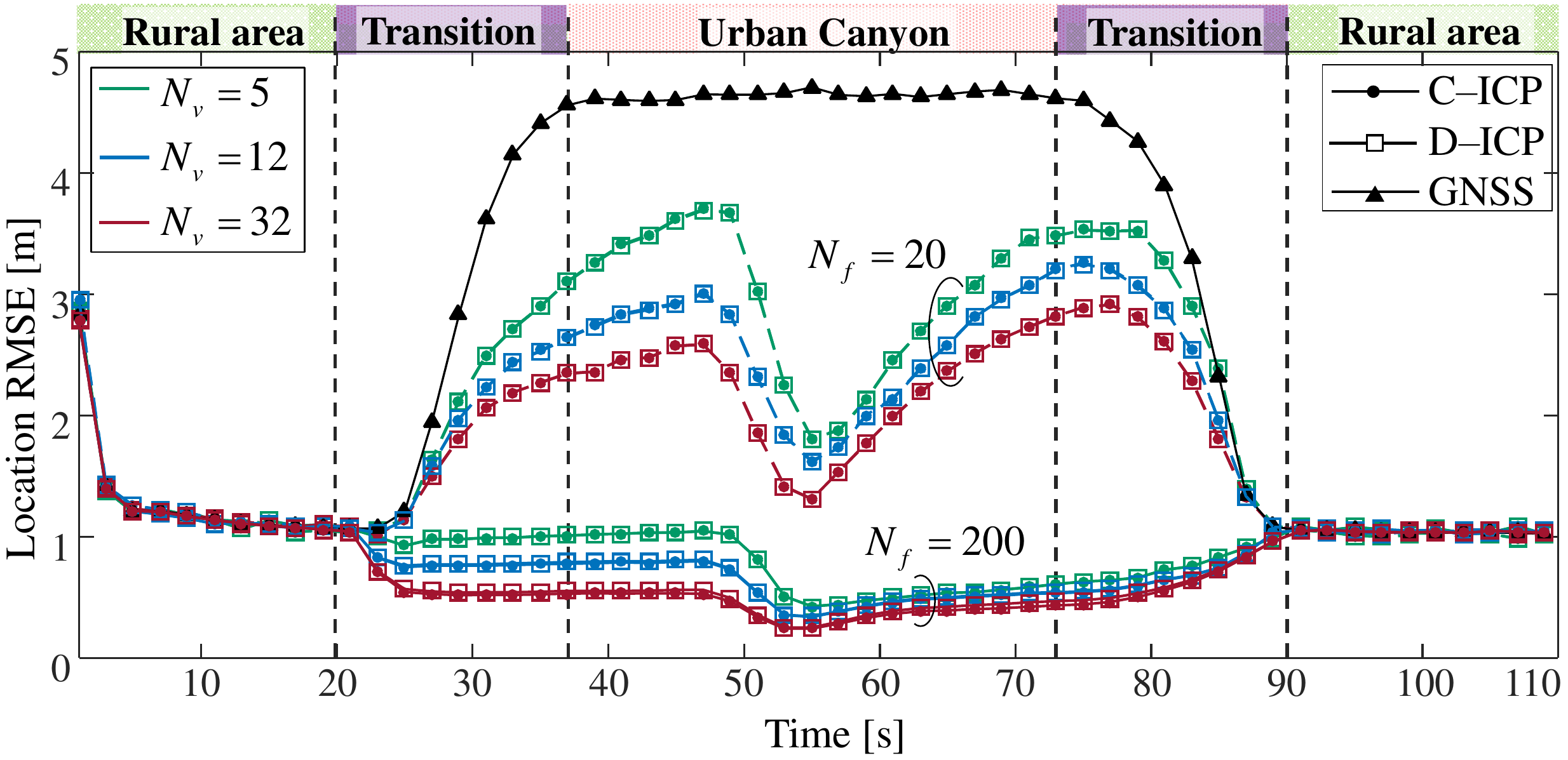}
\caption{RMSE of the vehicle position estimate versus time for the crossroad scenario in Fig. \ref{fig:results_fig4}, $N_v \in \{5, 12, 32\}$ vehicles and $N_f\in\{20,200\}$ features. The performance of the proposed distributed ICP algorithm is compared with both the stand-alone GNSS and centralized ICP approaches.}
\label{fig:results_fig5} 
\end{figure} 
We now evaluate, for $N_f \in \{ 20,200\}$, the impact of the number of vehicles. In Fig.~\ref{fig:results_fig5}, the RMSE of the vehicles' position estimate is shown versus time for  $N_v \in \{ 5,12,32 \}$ vehicles. It is clear that for a fixed number of features, more vehicles bring clear benefits in terms of positioning accuracy. In both Fig.~\ref{fig:results_fig4} and Fig.~\ref{fig:results_fig5}, we note an RMSE valley around $t = 55$ s. This is when most vehicles are in the urban canyon and there is high connectivity with many visible features.
 
\begin{figure}[!t]
\centering \includegraphics[width=1\columnwidth]{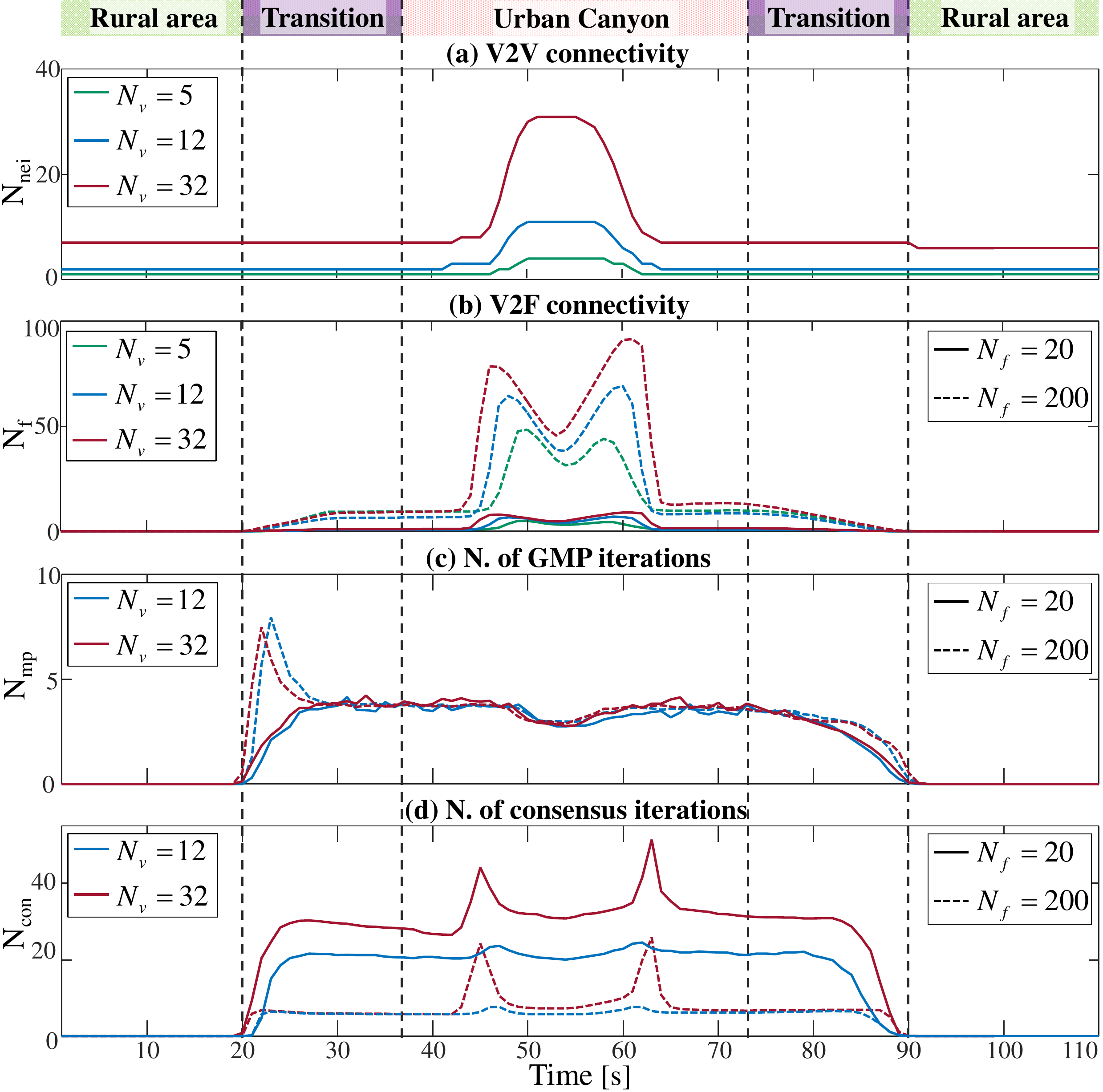}
\caption{Graph connectivity and number of iterations versus time for the ICP algorithm in the crossroad scenario of Fig. 4, with $N_v \in \{ 5,12,32\}$ vehicles and $N_f\in\{20,200\}$ features: (a) V2V connectivity, (b) V2F connectivity, (c) number of GMP iteration and (d) number of consensus iterations.}
\label{fig:results_fig6} 
\end{figure}

A detailed analysis of the connectivity is provided in Fig.~\ref{fig:results_fig6}. The V2V connectivity (i.e., average number of neighbors at each vehicle) is shown in Fig.~\ref{fig:results_fig6}-(a) and the V2F connectivity (i.e., average number of visible features at each vehicle) is in Fig.~\ref{fig:results_fig6}-(b), for a scenario with $N_f \in \{20,200\}$ features and $N_v \in \{ 5,12,32\}$ vehicles . We observe that since vehicles all start from the rural area and drive towards the urban canyon, there is a high V2V connectivity between $45$ s and $65$ s, which explains the behavior seen in Fig.~\ref{fig:results_fig5}. As highlighted at the bottom of Fig.~\ref{fig:results_fig4}, at the beginning and at the end of the observation time, there are four subgraphs (one per incoming road) since $R_c=150$ m and vehicles are $450$ m far from the intersection when they enter in the urban canyon area. On the other hand, in the proximity of the intersection all subgraphs merge into a single graph. The connectivity grows rapidly with the number of vehicles. For this connectivity to be useful, also the number of shared visible features needs to be sufficiently high. In Fig.~\ref{fig:results_fig6}-(b), we observe that this is again the case between $45$ s and $65$ s, due to the combination of two phenomena: a large number of connected vehicles and a large number of jointly observed features. Both are needed for the proposed algorithm to work well, as confirmed by Figs.~\ref{fig:results_fig4}--\ref{fig:results_fig5}. 

While high V2V and V2F connectivity are desirable, they come at a cost in delay. For the scenario with $N_v\in \{12,32\}$ vehicles and $N_f \in \{20,200\}$ features, Figs.~\ref{fig:results_fig6}-(c) and (d) illustrate the number of GMP iterations $N_{\mathrm{mp}}$ and consensus iterations $N_{\mathrm{con}}$ versus time, respectively. We observe that the number of GMP iterations rises rapidly when the vehicles enter the transition area, especially for a larger number of features, and remains roughly constant until they enter the second transition area. It is interesting to note that $N_{\mathrm{mp}}$ is relatively insensitive to the number of vehicles and features. While $N_{\mathrm{mp}}$ remains below 10 for all considered scenarios in Fig.~\ref{fig:results_fig6}-(c), $N_{\mathrm{con}}$ is generally larger (see Fig.~\ref{fig:results_fig6}-(d)). Moreover, it can be noticed that the number of consensus iterations increases around time instants $45$ s and $65$ s, i.e., respectively when the four subgraphs are fused into a single graph and when the single graph splits in four subgraphs, due to the low connectivity between vehicles at those time instants. In contrast to the GMP iterations, the number of consensus iterations increases with the number of vehicles, but decreases with the number of features. In fact, consensus convergence rate depends on the graph connectivity which is related to the number of features that connect single vehicles' subgraphs (see bold connections in Fig. \ref{fig:fact_graph}). The results from Fig.~\ref{fig:results_fig6} can be used to evaluate the communication overhead of the proposed distributed algorithm through (\ref{eq:comm_ov}).

\subsection{Real urban scenario with SUMO-simulated traffic}\label{sec:SUMO_Performance Evaluation}
\begin{figure}[!t]
	\centering
	\includegraphics[width=1\columnwidth]{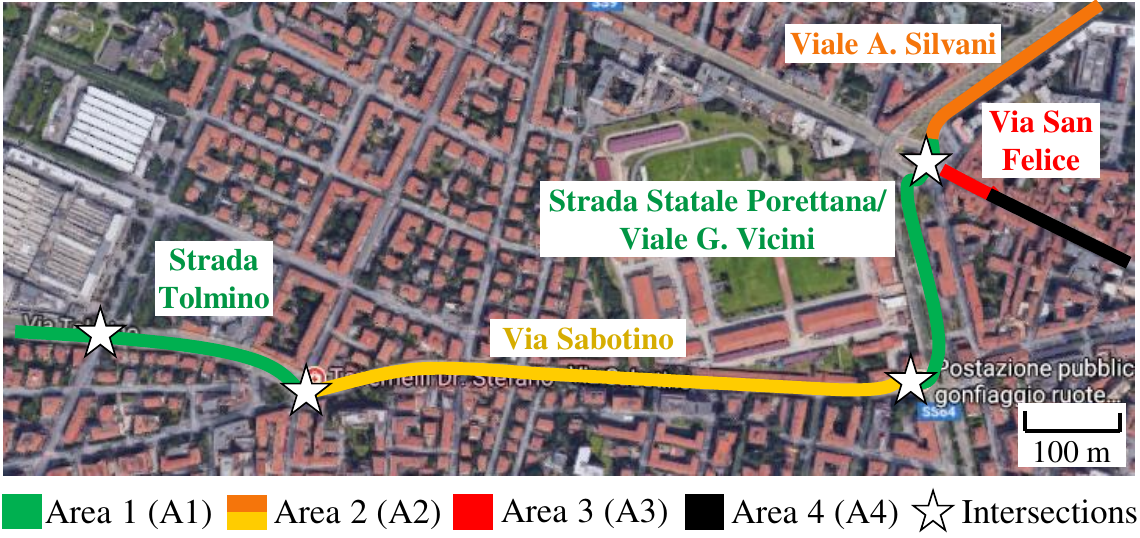}
	\caption{Map of the Bologna scenario, Italy, with vehicles and pedestrians are simulated over a \SI{1.2x0.5}{\kilo\meter} area. Each street is associated with a different GNSS signal quality (see Table \ref{tab:cases}).}
	\label{fig:Bologna}
\end{figure}
\textbf{\textit{Settings.}} To assess the ICP performance in a more realistic environment, we use the traffic simulator SUMO \cite{krajzewicz2012recent}, which uses real city maps to generate synthetic traces of vehicles and pedestrians. For this experiment, we consider vehicles and pedestrians, constrained to the highlighted streets, in a urban area of size \SI{1.2x0.5}{\kilo\meter} in the city of Bologna, Italy (see Fig.~\ref{fig:Bologna}). 
\begin{figure}
	\centering
	\includegraphics[width=1\columnwidth]{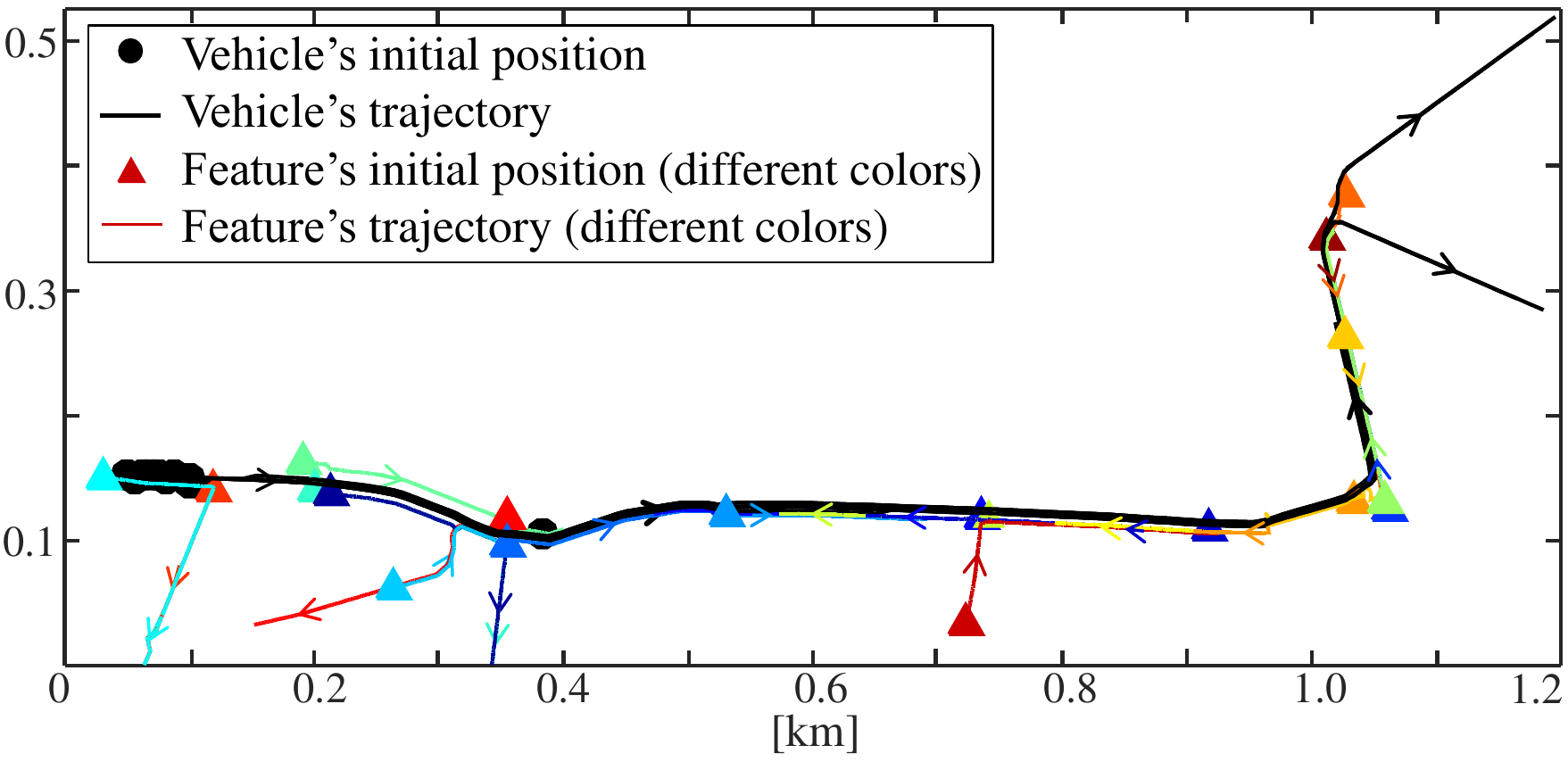}
	\caption{Superposition of 10 vehicle (thick black line) and 20 pedestrian (colored line) trajectories simulated by SUMO for $200$ s in the Bologna scenario of Fig. \ref{fig:Bologna}.}
	\label{fig:trajectories}
\end{figure}
In particular, we generate 10 vehicle (thick black line) and 20 pedestrian (colored line) trajectories (as shown in Fig.~\ref{fig:trajectories}) with sampling period $T_s=1$ s. The traces of the vehicles are synthesized according to a \textquotedblleft Krauss car-following\textquotedblright \ model, with maximum speed of $14$ m/s (around $50$ km/h), while the traces of the pedestrians are generated with an \textquotedblleft inter-trip chain\textquotedblright \ model which includes multi-modal profiles (e.g., purely static, queuing while entering a bus, walking on the sidewalk, suddenly turning to adjacent streets). The maximum pedestrian speed is set to $1.4$ m/s (about $5$ km/h) in our simulation. 

Each vehicle is assumed to equip a GNSS receiver and, in order to account for the wide diversity in the market, we assume four types of GNSS receivers \cite{denis2017cooperative}. Three vehicles are assigned a Standard Positioning Service (SPS) receiver whose position estimates have a standard deviation of $\bar{\sigma}_\text{GNSS}=\SI{3.6}{\meter}$, three other vehicles a Satellite-Based Augmentation Systems (SBAS) receiver with $\bar{\sigma}_\text{GNSS}=\SI{1.44}{\meter}$, two vehicles a Differential GNSS (DGNSS) receiver with $\bar{\sigma}_\text{GNSS}=\SI{40}{\centi\meter}$ and the last two vehicles a RTK receiver with $\bar{\sigma}_\text{GNSS}=\SI{1}{\centi\meter}$.
Moreover, since the GNSS accuracy is also sensitive to the surrounding environment, we model four types of environments which affect the quality of the GNSS differently, as shown in Table \ref{tab:cases}. 
\begin{table}[t!]
	\centering
	\caption{GNSS quality associated to each area of the Bologna's scenario in Fig. \ref{fig:Bologna} }
	\begin{tabular}{llll}
		\toprule
		Area & Street & Environment & \parbox[t]{2.2cm} {GNSS conditions} \\
		\midrule\rule{0pt}{2.5ex}
		A1 & \parbox[t]{1.8cm}{Via Tolmino\\Viale G. Vicini} & \parbox[t]{2.8cm}{Open sky, large road with 3 by 3 lanes, scattered med-size buildings} & \parbox[t]{2.2cm}{Nominal\\$\sigma_\text{GNSS}=1\bar{\sigma}_\text{GNSS}$}\\
		\rule{0pt}{2.5ex}
		A2 & \parbox[t]{1.8cm}{Via Sabotino\\Viale A. Silvani}  & \parbox[t]{2.8cm}{Some blockage, narrow road, 3 lanes, scattered medium-size buildings} & \parbox[t]{2.2cm}{Slightly degraded \\ $\sigma_\text{GNSS}=2\bar{\sigma}_\text{GNSS}$} \\
		\rule{0pt}{2.5ex}
		A3 & \parbox[t]{1.8cm}{Via San Felice}  & \parbox[t]{2.8cm}{Ultra narrow road, \\2 lanes,	urban canyon} & \parbox[t]{2.2cm}{Severely degraded\\ $\sigma_\text{GNSS}=5 \bar{\sigma}_\text{GNSS}$} \\
		\rule{0pt}{2.5ex}
		A4 & \parbox[t]{1.8cm}{Via San Felice}  & \parbox[t]{2.8cm}{Ultra narrow road, \\2 lanes,	urban canyon} & \parbox[t]{2.2cm}{Lost\\$\sigma_\text{GNSS}=20\bar{\sigma}_\text{GNSS}$ }\\
		\bottomrule
	\end{tabular}
	\label{tab:cases}
\end{table}
The third column of the table indicates how much the standard deviation of GNSS measurements is incremented with respect to their nominal value $\bar{\sigma}_\text{GNSS}$. The simulated traces are used to determine the ground-truth reference, to calibrate vehicle/feature mobility models and to produce synthetic erroneous measurements. Tracking is performed by using the mobility models of vehicles and features respectively in (\ref{eq:veh_motion_model}) and (\ref{eq:fea_motion_model}). The standard deviation of V2F sensing is set to $\sigma_{\mathrm{V2F}} = 0.1$ m (as representative of RADAR accuracy \cite{denis2017cooperative}). The communication range at each vehicle is set to $R_c = 200$ m, while the sensing range $R_s$ is assumed to be lower and varies through simulations.

\begin{figure}
	\centering
	\includegraphics[width=1\columnwidth]{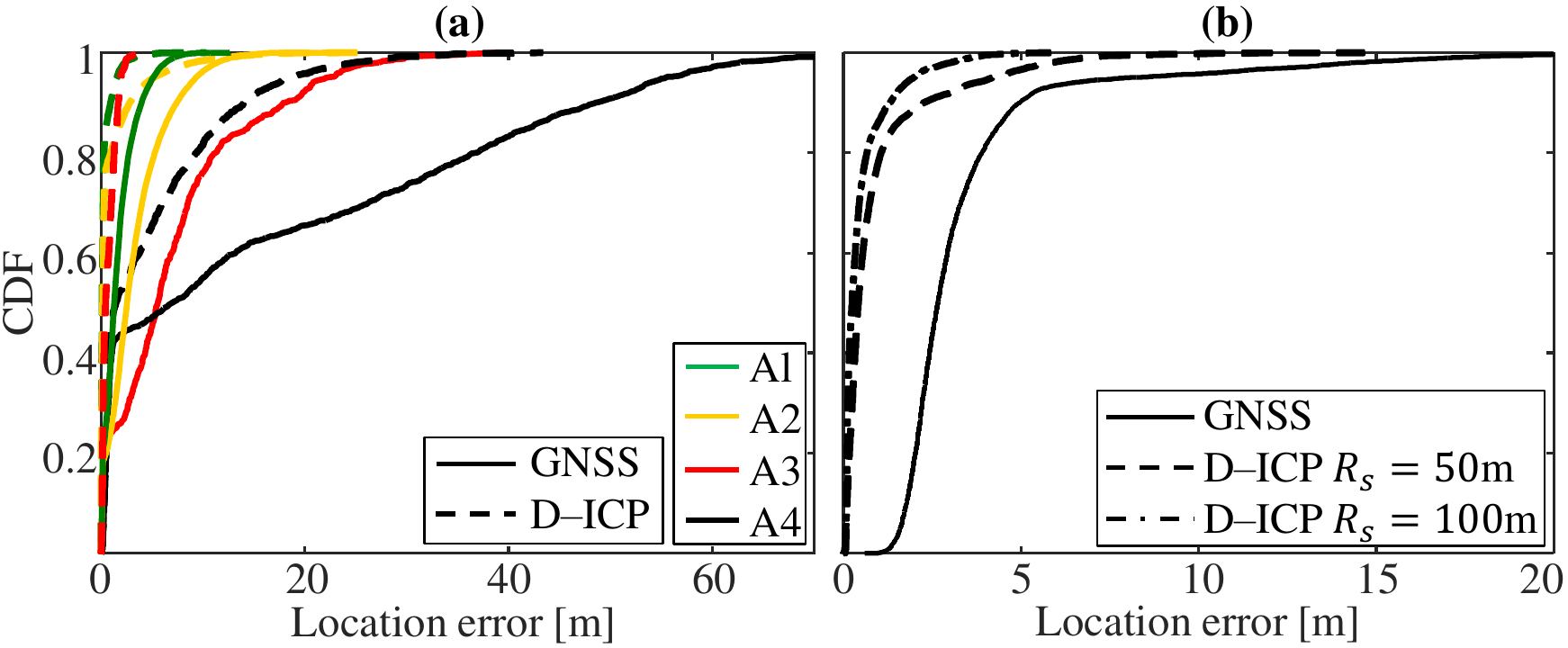}
	\caption{CDF of the vehicle location error for the Bologna scenario in Fig. \ref{fig:Bologna}, for the distributed ICP and stand-alone GNSS. Positioning accuracy (a) over different areas for $R_s = 50$ m and (b) for $R_s = 50$ m and $R_s = 100$ m.}
	\label{fig:CDF}
\end{figure}

\textbf{\textit{Numerical results.}} In Fig. \ref{fig:CDF}-(a), performances in terms of CDF of the location error are illustrated for different GNSS qualities associated to each street/area (see Table \ref{tab:cases}). As expected, performance improvements are observed when the distributed ICP (dashed line) method is used with respect to the stand-alone GNSS (solid line) one. Fig. \ref{fig:CDF}-(b) shows the CDF of the vehicle location error for the distributed ICP method with different sensing ranges, $R_s = 50$ m and $R_s = 100$ m (respectively, dashed and dashed-dot lines), compared with the stand-alone GNSS (solid line). For $50\%$ of confidence level, the ICP approach achieves a location accuracy of $0.46$ m for $R_s = 50$ m and $0.23$ m for $R_s = 100$, while the stand-alone GNSS accuracy is $2.65$ m. 

\begin{figure}
	\centering
	\includegraphics[width=1\columnwidth]{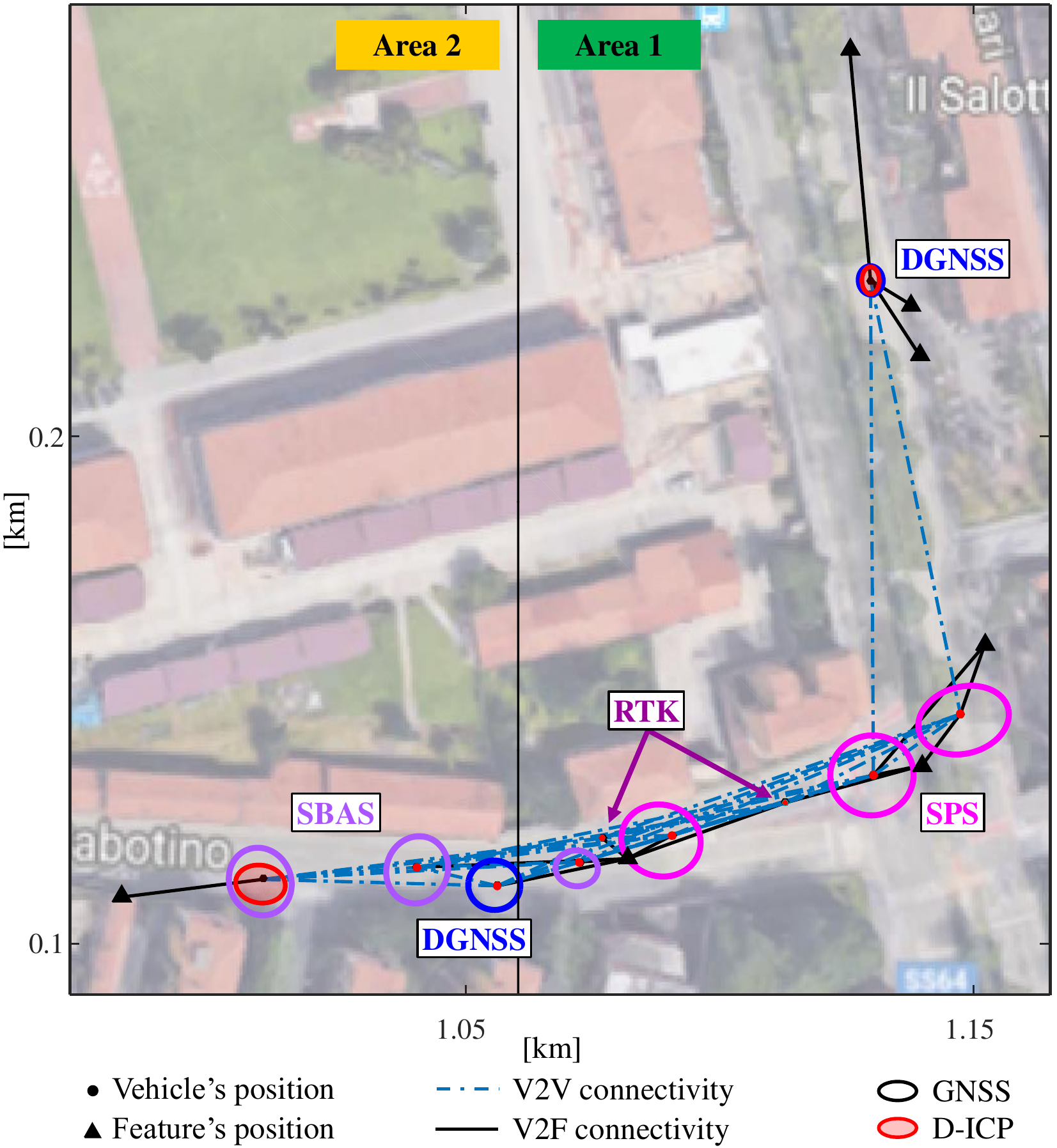}
	\caption{Localization accuracy for the Bologna scenario of Fig. \ref{fig:Bologna} over areas 1 and 2 at time $t = 137$ s, for sensing ranges $R_s = 50$ m: distributed ICP (red ellipse) and stand-alone GNSS (colored contours based on receiver type). Zoomed view over the intersection between Viale Sabotino (A2) and Viale G. Vicini (A1).}
	\label{fig:ellipse_sumo}
\end{figure}
Zoomed view of Bologna city over the intersection between Viale Sabotino (area 2) and Viale G. Vicini (area 1) is shown in Fig. \ref{fig:ellipse_sumo}. Here, the average performances are evaluated by computing the $2 \times 2$ mean square error matrix of vehicles' position estimates at convergence, $\mathbf{MSE} = E[(\hat{\mathbf{p}}^{(\mathrm{V})}_{i,t}- \mathbf{p}^{(\mathrm{V})}_{i,t})(\hat{\mathbf{p}}^{(\mathrm{V})}_{i,t}- \mathbf{p}^{(\mathrm{V})}_{i,t})^\mathrm{T}]$, over 100 independent observations, for both the distributed ICP algorithm (red ellipse) and stand-alone GNSS method for different types of GNSS receivers (coloured contours). For visualization purposes, the error ellipses at $98.9$\% confidence are plotted around the mean vehicles' position estimates. The V2F and V2V connectivities are also given (respectively black solid and grey dashed-dot lines). Results show that all vehicles improve their position accuracy by using the ICP method compared to the performances obtained by the stand-alone GNSS solution. Note that the location accuracy given by the proposed ICP algorithm is not uniform among vehicles as it depends on the type of GNSS receiver, on the GNSS signal quality in the area in which vehicles are traveling, but also on the vehicles' and features' positions.  
%
%Fig. \ref{fig:cdf_gpskind} shows the cumulative distribution function (CDF) of the vehicle location error. The proposed distributed ICP method (red dashed line) is compared with the centralized ICP (blue solid line) and the stand-alone GNSS (black solid line) solutions, for sensing range $R_s = 100$ m and for different GNSS receivers: Figs. \ref{fig:cdf_gpskind}-(a) SPS, (b) SBAS, (c) DGNSS and (d) RTK (with cm-level accuracy). Results show that the distributed ICP algorithm achieves sub-meter accuracy for $75$\% of confidence level in all four cases and outperforms the stand-alone GNSS solution, even when RTK receiver is used though with a small gain. Moreover, the proposed distributed method attains the centralized approach.  
%\begin{figure}[t!]
%	\centering
%	\includegraphics[width=1\columnwidth]{cdf_gpskind_v5}
%	\caption{CDF of the vehicle location error for the Bologna's scenario in Fig. \ref{fig:Bologna}, for the distributed (D-ICP) and centralized (C-ICP) cooperative methods and the stand-alone GNSS solution with four types of GNSS receivers: (a) SPS for vehicles $i=1,5,9$, (b) SBAS for vehicles $i=2,6,10$, (c) DGNSS for vehicles $i=3,7$ and (d) RTK for vehicles $i=4,8$.}
%	\label{fig:cdf_gpskind}
%\end{figure}

\section{Conclusion}\label{sec:Conclusion}

In this paper, a novel framework of cooperative positioning in vehicular networks was proposed, in which vehicles had to estimate a set of common passive features in a fully distributed way to improve the GNSS-based vehicle positioning. Starting from a FG formulation of the positioning problem, we developed a distributed Gaussian message passing algorithm that employed a consensus-based scheme for the distributed estimation of the features' positions. Simulation results demonstrated that the proposed methodology can accurately estimate the features' positions and (implicitly) improve the vehicle positioning accuracy compared to the stand-alone GNSS solution. Moreover, the ICP method was validated in a real urban scenario using the SUMO traffic simulator. 

The framework made several limiting assumptions. First of all, the assumption of a linear measurement model can be removed by considering arbitrary non-linear models with non-parametric (e.g., particles) or parametric (e.g., Gaussian mixtures) message representations. Secondly, the assumption of perfect data association can be removed by including the data association problem in the FG. Investigation of these issues is a topic of further research.

\section*{Appendix}\label{sec:Appendix}

Based on the result in \eqref{eq:veh_sigma_final} and recalling that $\alpha_{p}^{\textsc{(V)}}=N_{f}/\sigma_{\mathrm{V2F}}^{2}+1/\sigma_{p,\text{pr}}^{\textsc{(V)}^2}+1/\sigma_{\text{GNSS}}^{2}$, the limiting cases considered in Sect. \ref{sec:ICP Performance Analysis} are derived as follows:
\begin{enumerate}
\item If $N_v \rightarrow \infty$, \eqref{eq:high_Nv} is given by:
\begin{equation}
\small
\sigma_{p,\text{post}}^{{\textsc{(V)}}^2} \rightarrow \frac{1}{\alpha_{p}^{\textsc{(V)}}}\cdot \frac{N_v/\sigma_{p,\text{pr}}^{\textsc{(V)}^2} + N_v/\sigma_{\text{GNSS}}^2}{N_v/\sigma_{p,\text{pr}}^{\textsc{(V)}^2} + N_v/\sigma_{\text{GNSS}}^2}= \frac{1}{\alpha_{p}^{\textsc{(V)}}}.   
\end{equation}

\item If $N_f \rightarrow \infty$, \eqref{eq:high_Nf} is obtained as:
\begin{equation}
\small
\hspace{-0.4cm}\sigma_{p,\text{post}}^{{\textsc{(V)}}^2}\rightarrow \frac{1}{N_f/\sigma_{\mathrm{V2F}}^2}\cdot \frac{N_f/\sigma_{\mathrm{V2F}}^2 \hspace{-0.04cm}+\hspace{-0.04cm} N_f/\sigma_{p,\text{pr}}^{\textsc{(F)}^2}}{N_f/\sigma_{p,\text{pr}}^{\textsc{(F)}^2}}
\rightarrow\frac{\sigma_{p,\text{pr}}^{\textsc{(F)}^2}\hspace{-0.04cm}+\hspace{-0.04cm} \sigma_{\mathrm{V2F}}^2}{N_f}\rightarrow 0.      
\end{equation}
%\begin{split}
%\sigma_{p,\text{post}}^{{\textsc{(V)}}^2} &\rightarrow \frac{1}{N_f/\sigma_{\mathrm{V2F}}^2}\cdot \frac{N_f/\sigma_{\mathrm{V2F}}^2 + N_f/\sigma_{p,\text{pr}}^{\textsc{(F)}^2}}{N_f/\sigma_{p,\text{pr}}^{\textsc{(F)}^2}}\\
%&\rightarrow\frac{\sigma_{p,\text{pr}}^{\textsc{(F)}^2} + \sigma_{\mathrm{V2F}}^2}{N_f}\rightarrow 0.      
%\end{split}
\item If $\sigma_{\mathrm{V2F}}^2\rightarrow 0$, \eqref{eq:low_sigma_r} is:
\begin{equation}
\small
\begin{split}
\sigma_{p,\text{post}}^{{\textsc{(V)}}^2} & \rightarrow \frac{1}{N_f/\sigma_{\mathrm{V2F}}^2}\cdot \frac{N_f/\sigma_{\mathrm{V2F}}^2}{N_v/\sigma_{p,\text{pr}}^{\textsc{(V)}^2} + N_v/\sigma_{\text{GNSS}}^2+ N_f/\sigma_{p,\text{pr}}^{\textsc{(F)}^2}} \\
& = \frac{1}{N_v/\sigma_{p,\text{pr}}^{\textsc{(V)}^2} + N_v/\sigma_{\text{GNSS}}^2+ N_f/\sigma_{p,\text{pr}}^{\textsc{(F)}^2}}.
\end{split}        
\end{equation}
    
\item If $\sigma_{p,\text{pr}}^{\textsc{(F)}^2}\rightarrow 0$, \eqref{eq:low_sigma_f} is given as:
\begin{equation}
\small
\begin{split}
\hspace{-0.4cm}\sigma_{p,\text{post}}^{{\textsc{(V)}}^2} &\rightarrow\hspace{-0.05cm} \frac{1}{\alpha_{p}^{\textsc{(V)}}}\hspace{-0.07cm}\cdot\hspace{-0.07cm}\frac{N_f/\sigma_{p,\text{pr}}^{\textsc{(F)}^2} \hspace{-0.09cm}+\hspace{-0.07cm} \sigma_{\mathrm{V2F}}^2/\hspace{-0.03cm}(\sigma_{p,\text{pr}}^{\textsc{(V)}^2}\sigma_{p,\text{pr}}^{\textsc{(F)}^2}\hspace{-0.02cm})\hspace{-0.09cm}+\hspace{-0.07cm} \sigma_{\mathrm{V2F}}^2/\hspace{-0.03cm}(\sigma_{\text{GNSS}}^2\sigma_{p,\text{pr}}^{\textsc{(F)}^2}\hspace{-0.02cm})}{N_f/\sigma_{p,\text{pr}}^{\textsc{(F)}^2} \hspace{-0.09cm}+\hspace{-0.07cm} \sigma_{\mathrm{V2F}}^2/\hspace{-0.03cm}(\sigma_{p,\text{pr}}^{\textsc{(V)}^2}\sigma_{p,\text{pr}}^{\textsc{(F)}^2}\hspace{-0.02cm})\hspace{-0.09cm}+\hspace{-0.07cm} \sigma_{\mathrm{V2F}}^2/\hspace{-0.03cm}(\sigma_{\text{GNSS}}^2\sigma_{p,\text{pr}}^{\textsc{(F)}^2}\hspace{-0.03cm})} \\&= \frac{1}{\alpha_{p}^{\textsc{(V)}}}.
\end{split} 
\end{equation}

\item If $\sigma_{p,\text{pr}}^{\textsc{(V)}^2}\rightarrow 0$ or if $\sigma_{\text{GNSS}}^2\rightarrow 0$, then \eqref{eq:low_sigma_v_or_GNSS} is obtained as: 
\begin{equation}
\small
\begin{split}
\hspace{-0.4cm}\sigma_{p,\text{post}}^{{\textsc{(V)}}^2} &\rightarrow \frac{1}{1/\sigma_{\text{GNSS}}^2 + 1/\sigma_{p,\text{pr}}^{\textsc{(V)}^2}} \cdot \\
& \hspace{-0.45cm} \frac{N_v/\sigma_{p,\text{pr}}^{\textsc{(V)}^2}\hspace{-0.08cm}+\hspace{-0.06cm} N_v/\sigma_{\text{GNSS}}^2 \hspace{-0.09cm}+\hspace{-0.07cm}\sigma_{\mathrm{V2F}}^2/(\sigma_{p,\text{pr}}^{\textsc{(V)}^2}\sigma_{p,\text{pr}}^{\textsc{(F)}^2}\hspace{-0.02cm})\hspace{-0.08cm}+\hspace{-0.06cm} \sigma_{\mathrm{V2F}}^2/(\sigma_{\text{GNSS}}^2\sigma_{p,\text{pr}}^{\textsc{(F)}^2}\hspace{-0.02cm})}{N_v/\sigma_{p,\text{pr}}^{\textsc{(V)}^2}\hspace{-0.09cm}+\hspace{-0.07cm} N_v/\sigma_{\text{GNSS}}^2 \hspace{-0.09cm}+\hspace{-0.07cm} \sigma_{\mathrm{V2F}}^2/(\sigma_{p,\text{pr}}^{\textsc{(V)}^2}\sigma_{p,\text{pr}}^{\textsc{(F)}^2}\hspace{-0.02cm})\hspace{-0.09cm}+\hspace{-0.07cm} \sigma_{\mathrm{V2F}}^2/(\sigma_{\text{GNSS}}^2\sigma_{p,\text{pr}}^{\textsc{(F)}^2}\hspace{-0.02cm})}\\
& \hspace{-0.45cm} = \frac{1}{1/\sigma_{\text{GNSS}}^2 + 1/\sigma_{p,\text{pr}}^{\textsc{(V)}^2}}\rightarrow 0.
\end{split}        
\end{equation}

\end{enumerate}

\bibliographystyle{IEEEtran}
\bibliography{Bib_ICP_v4}

\begin{IEEEbiography}[{\includegraphics[width=1in,height=1.25in,clip,keepaspectratio]{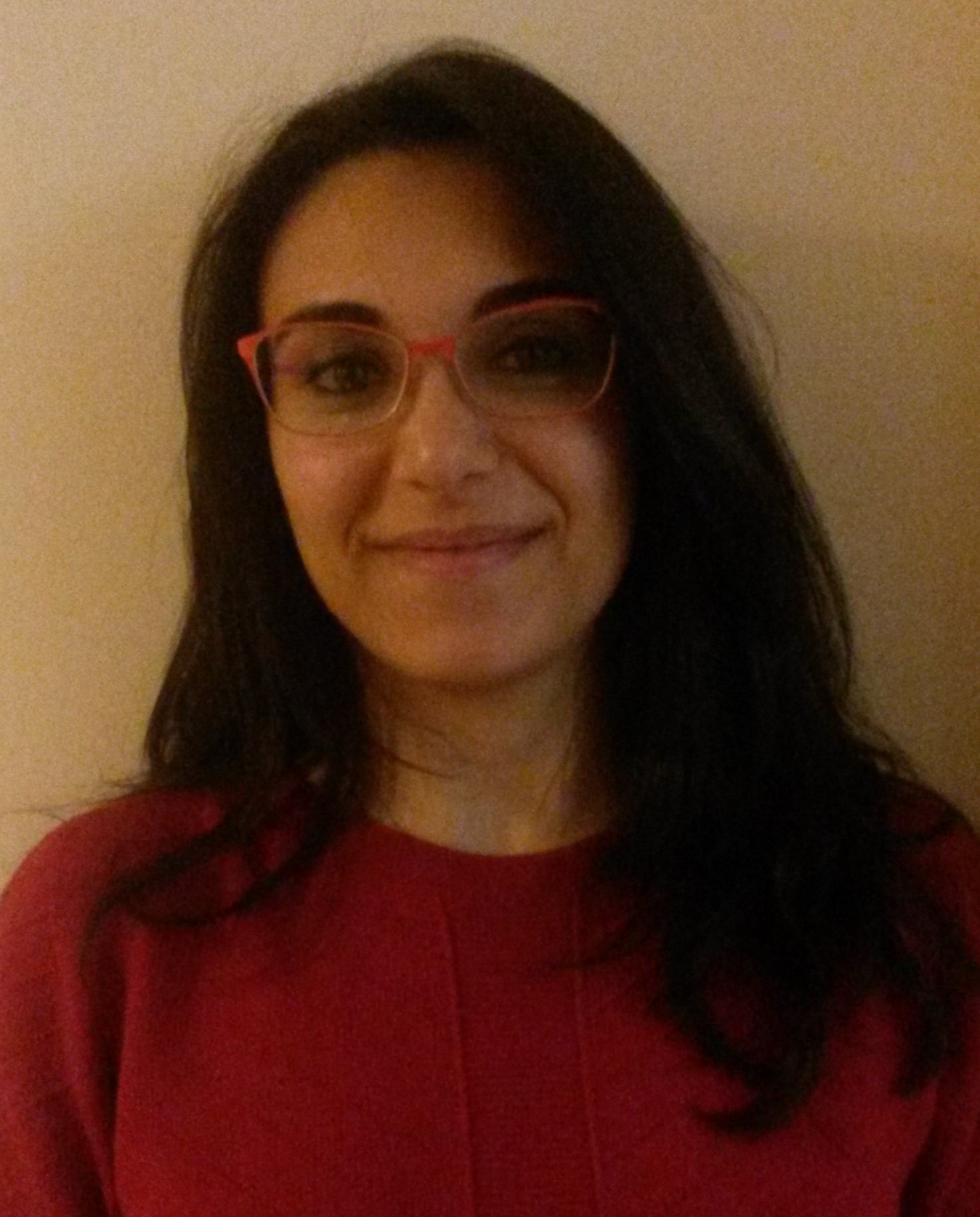}}]{Gloria Soatti} received the M.Sc. degree in Telecommunication Engineering in 2012 and the Ph.D. degree (cum laude) in Information Technology in Feb. 2017 both from Politecnico di Milano (Italy). She was a visiting researcher at the Department of Signals and Systems, Chalmers University of Technology (Sweden) in 2016. Currently, she is a post-doctoral researcher at the Dipartimento di Elettronica, Informazione e Bioingegneria (DEIB) of Politecnico di Milano. Her research interests are in the field of signal processing, particularly distributed consensus-based approaches for wireless sensor networks, IoT cognitive radios and vehicular networks.
\end{IEEEbiography}

\begin{IEEEbiography}[{\includegraphics[width=1in,height=1.25in,clip,keepaspectratio]{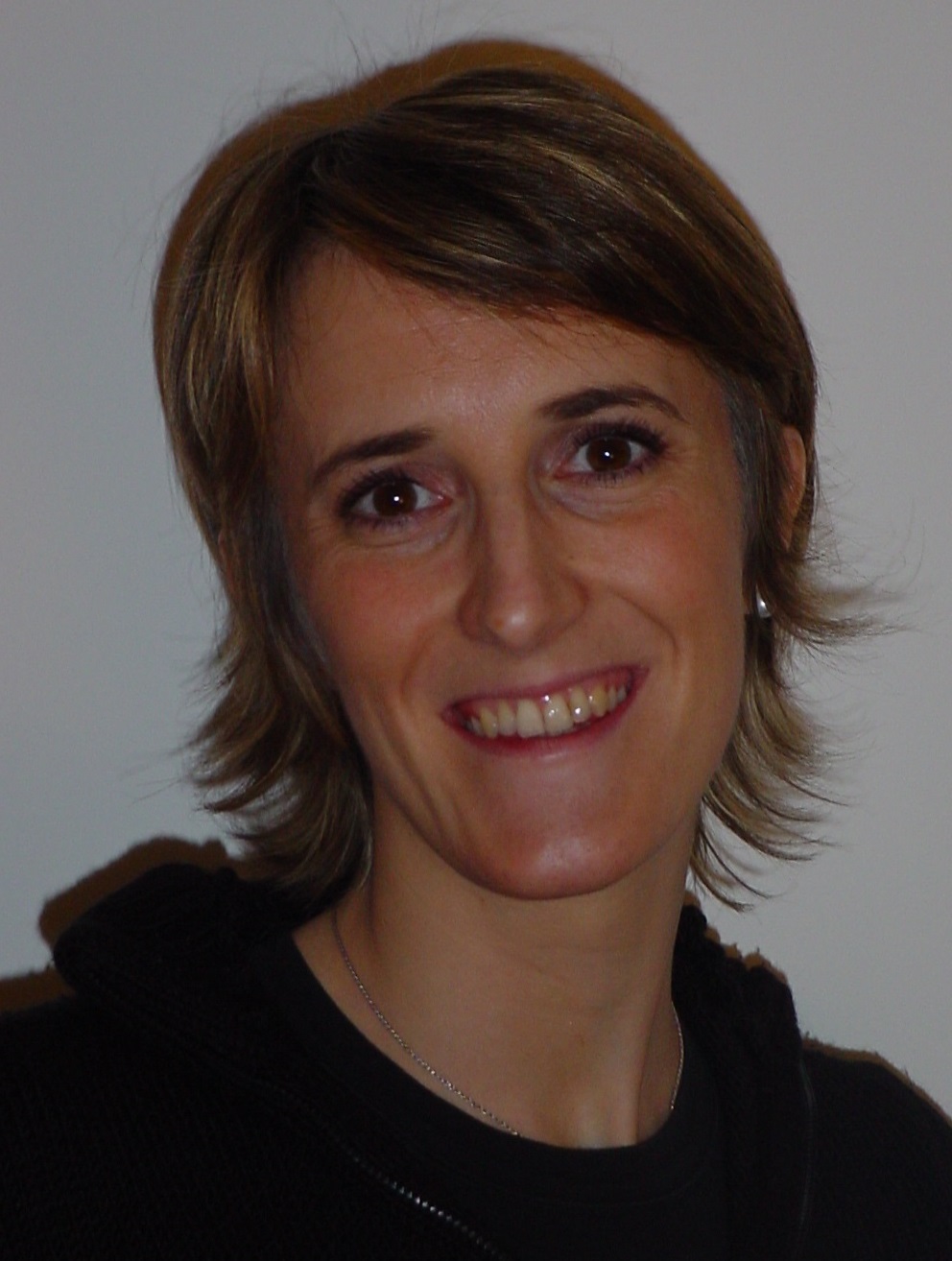}}]{Monica Nicoli} (M’99) received the M.Sc. degree (with honors) and the Ph.D. degree in Telecommunication Engineering from Politecnico di Milano, in 1998 and 2002, respectively. During 2001 she was Visiting Researcher with Uppsala University, Sweden. Since 2002 she has been with the Dipartimento di Elettronica, Informazione e Bioingegneria, Politecnico di Milano, where she is Assistant Professor. Her research interests are in the area of signal processing, with emphasis on wireless communications, distributed and cooperative systems, intelligent transportation systems, wireless positioning and navigation. Dr. Nicoli is an Associate Editor of the EURASIP Journal on Wireless Communications and Networking, and she served as Lead Guest Editor for the Special Issue on “Localization in Mobile Wireless and Sensor Networks” in 2011. She has been member of the technical program committees of several conferences in the area of signal processing and wireless communications. She received the Marisa Bellisario Award in 1999 and the Premium Award for the Best Paper in IET Intelligent Transport Systems in 2014.
\end{IEEEbiography}

\begin{IEEEbiography}[{\includegraphics[width=1in,height=1.25in,clip,keepaspectratio]{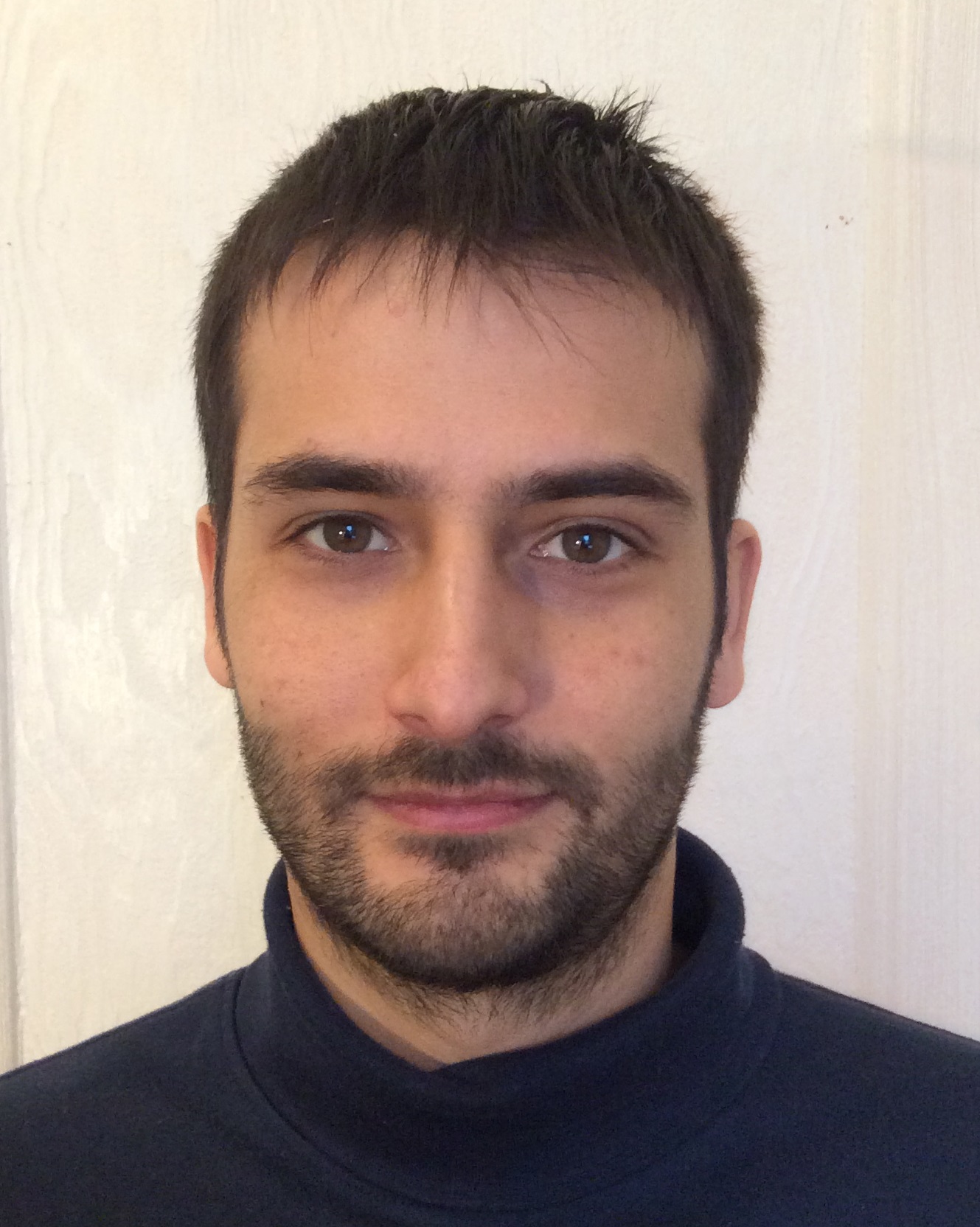}}]{Nil Garcia} (S’14, M’16) received the Telecommunications Engineer degree from the Polytechnic University of Catalonia (UPC), Barcelona, Spain, in 2008; and the double Ph.D. degree in electrical engineering from the New Jersey Institute of Technology, Newark, NJ, USA, and from the National Polytechnic Institute of Toulouse, Toulouse, France, in 2015. He is currently a postdoctoral researcher of Communication Systems with the Department of Signals and Systems at Chalmers University of Technology, Sweden. In 2009, he worked as an engineer in the Centre National d'\'{E}tudes Spatiales (CNES). In 2008 and 2009, he had Internships in CNES and NASA. His research interests are in the areas of localization, intelligent transportation systems and 5G.
\end{IEEEbiography}

\begin{IEEEbiography}[{\includegraphics[width=1in,height=1.25in,clip,keepaspectratio]{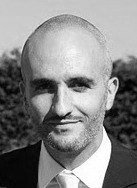}}]{Benoit Denis} received the E.E. (2002), M.Sc. (2002), and Ph.D. (2005) degrees from INSA (Rennes, France) in electronics and communication systems. Since December 2005, he has been with CEA-Leti Minatec (Grenoble, France), contributing into French (ANR), European (FP6/FP7/H2020) and extra-European (QNRF) collaborative research projects in the fields of wireless sensor networks and wearable networks, heterogeneous and cooperative networks, vehicular networks, and mobile applications related to the Internet of Things or Connected Intelligent Transportation Systems. His main research interests concern joint wireless localization and communications, ranging/positioning/tracking and hybrid data fusion algorithms, radio channel modeling and cross-layer protocol design. He as (co)authored about 100 scientific papers on the previous topics.
\end{IEEEbiography}

\begin{IEEEbiography}[{\includegraphics[width=1in,height=1.25in,clip,keepaspectratio]{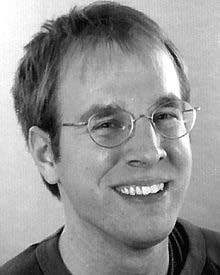}}]{Ronald Raulefs} received the Dipl.-Ing. degree from the University of Kaiserslautern, Germany, in 1999 and the Dr.-Ing. (PhD) degree from the University of Erlangen-Nuremberg, Germany, in 2008. He is working as senior research member at the Institute of Communications and Navigation of the German Aerospace Center (DLR) in Oberpfaffenhofen, Germany. Ronald Raulefs initiated and lead the EU FP7 project WHERE and its successor project WHERE2 (www.ict-where2.eu) as well as the task on cooperative location and communications in heterogeneous networks. He taught courses on the cooperation between wireless communications and positioning systems, such as the tutorials at the VTC'09, Sarnoff Symposium (2010), Summer school of WHERE/WHERE2 (2010), European Wireless (2013), ICC'13, Winter school Newcom\#/IC 1004 (2013) and ICC'17. He authored and co-authored 80+ scientific publications in conferences and journals. His current research interests include various aspects of mobile radio communications and positioning, including cooperative positioning for future cellular communication systems.
\end{IEEEbiography}

\begin{IEEEbiography}[{\includegraphics[width=1in,height=1.25in,clip,keepaspectratio]{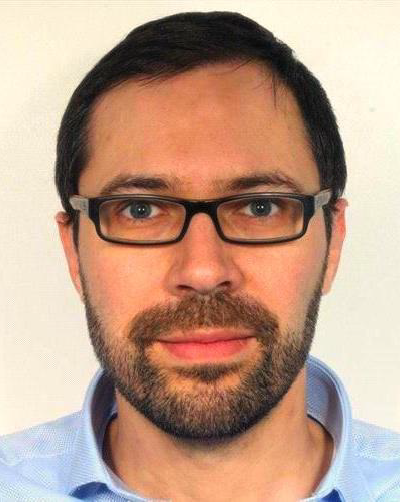}}]{Henk Wymeersch} (S’01, M’05) obtained the Ph.D. degree in Electrical Engineering/Applied Sciences in 2005 from Ghent University, Belgium. He is currently a Professor of Communication Systems with the Department of Signals and Systems at Chalmers University of Technology, Sweden. Prior to joining Chalmers, he was a postdoctoral researcher from 2005 until 2009 with the Laboratory for Information and Decision Systems at the Massachusetts Institute of Technology. Prof. Wymeersch served as Associate Editor for IEEE Communication Letters (2009-2013), IEEE Transactions on Wireless Communications (since 2013), and IEEE Transactions on Communications (since 2016). His current research interests include cooperative systems and intelligent transportation. 
\end{IEEEbiography}

\end{document}